\documentclass[journal=jacsat,manuscript=communication]{achemso}
\setkeys{acs}{articletitle = true}

\usepackage{chemformula} 
\usepackage[T1]{fontenc} 
\usepackage{graphicx}
\usepackage{pdfpages}
\usepackage{amsmath}
\usepackage{pdfpages}

\author{Doran I. G. Bennett}
\email{doranbennett@g.harvard.edu}
\affiliation{Department of Chemistry and Chemical Biology, Harvard University, 12 Oxford St, Cambridge, Massachusetts 02138, USA}
\alsoaffiliation{Bio-Inspired Solar Energy Program, Canadian Institute for Advanced Research, Toronto, Ontario M5G 1Z8, Canada}

\author{Pavel Mal\'{y}}
\affiliation{Department of Physics and Astronomy, Faculty of Sciences, VU University Amsterdam, De Boelelaan 1081, 1081 HV Amsterdam, The Netherlands}
\alsoaffiliation{Institute of Physics, Faculty of Mathematics and Physics, Charles University, Ke Karlovu 3, 121 16 Prague 2, Czech Republic}
\alsoaffiliation{Bio-Inspired Solar Energy Program, Canadian Institute for Advanced Research, Toronto, Ontario M5G 1Z8, Canada}

\author{Christoph Kreisbeck}
\affiliation{Department of Chemistry and Chemical Biology, Harvard University, 12 Oxford St, Cambridge, Massachusetts 02138, USA}

\author{Rienk van Grondelle}
\affiliation{Department of Physics and Astronomy, Faculty of Sciences, VU University Amsterdam, De Boelelaan 1081, 1081 HV Amsterdam, The Netherlands}
\alsoaffiliation{Bio-Inspired Solar Energy Program, Canadian Institute for Advanced Research, Toronto, Ontario M5G 1Z8, Canada}

\author{Al\'{a}n Aspuru-Guzik}
\affiliation{Department of Chemistry and Chemical Biology, Harvard University, 12 Oxford St, Cambridge, Massachusetts 02138, USA}
\alsoaffiliation{Bio-Inspired Solar Energy Program, Canadian Institute for Advanced Research, Toronto, Ontario M5G 1Z8, Canada}

\title{Mechanistic Regimes of Vibronic Transport in a Heterodimer and the Design Principle of Incoherent Vibronic Transport in Phycobiliproteins}

\abbreviations{}

\begin{document}


\begin{abstract}
	Following the observation of coherent oscillations in non-linear spectra of photosynthetic pigment protein complexes, particularly phycobilliprotein such as PC645, coherent vibronic transport has been suggested as a design principle for novel light harvesting materials operating at room temperature. Vibronic transport between energetically remote pigments is coherent when the presence of a resonant vibration supports transient delocalization between the pair of electronic excited states. Here, we establish the mechanism of vibronic transport for a model heterodimer across a wide range of molecular parameter values. The resulting mechanistic map demonstrates that the molecular parameters of phycobiliproteins in fact support incoherent vibronic transport. This result points to an important design principle: incoherent vibronic transport is more efficient than a coherent mechanism when energetic disorder exceeds the coupling between the donor and vibrationally excited acceptor states. Finally, our results suggest that the role of coherent vibronic transport in pigment protein complexes should be reevaluated. 
\end{abstract}

Excitation transport down an energy gradient, like that observed in some photosynthetic or artificial light harvesting complexes, requires the dissipation of excess electronic energy into molecular vibrations. Vibronic transport is a photophysical process that converts an electronic excitation on one pigment to an electronic and vibrational excitation on another pigment (or vice-versa). Vibronic transport between detuned pigments has been identified as a potential design principle for accelerating or controlling exciton migration in next generation materials \cite{Scholes2017a,OReilly2014a,Dean2016a,Killoran2015a,Blau2017a}. Realizing the advantages of engineered vibrational environments in practical devices, however, requires a clear understanding of how the mechanism of vibronic  transport changes as a function of the chemical structure and vibrational dynamics of the pigments.  

Vibronic transport is coherent when the resonant, high-frequency vibration supports transient delocalization between energetically remote pigments. Coherent vibronic transport allows for a ballistic spread of excitation density which outraces the diffusive transport supported by incoherent mechanisms. Following the observation of coherent oscillations in non-linear spectroscopy, many researchers have suggested biological pigment protein complexes (PPC) use coherent vibronic transport to enhance the rate of light harvesting \cite{Chin2013a,Romero2014a,Fuller2014a,Nalbach2015a,Dean2016a}, but this remains controversial \cite{Fujihashi2015a,Blau2017a, Duan02017a}. Here, we will simulate a minimal vibronic heterodimer to establish how the transport mechanism varies across molecular parameter space. These results map out the mechanistic regimes and are appropriate for analyzing a wide variety of vibronic dimers. We use the mechanistic map to demonstrate that phycobiliproteins, a family of PPCs that remains a cannonical example of coherent vibronic transport thought to occur at room temperature \cite{Womick2011a,Kolli2012a,  Dean2016a,Scholes2017a,OReilly2014a}, in fact undergo incoherent vibronic transport. This result points to a basic design principle we suggest is important for understanding vibrationally mediated exciton transport in both natural and artificial materials: incoherent vibronic transport is more efficient than a coherent mechanism when energetic disorder exceeds the coupling between the donor and vibrationally excited acceptor states. Finally, our results suggest that the extent of coherence in vibronic transport for non-bilin PPCs should also be reevaluated. 

\begin{figure*}[t!]
	\begin{center}
		\includegraphics{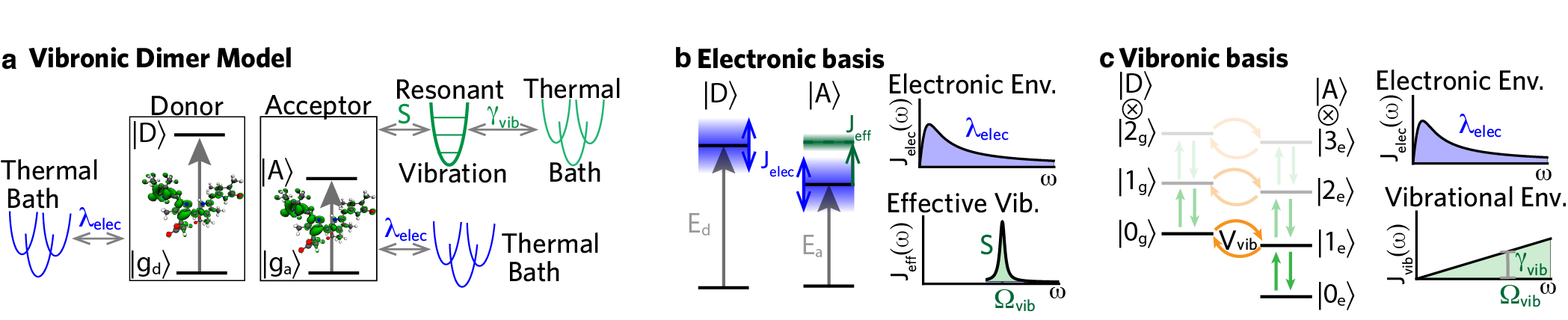}
		\caption{(a) Schematic representation of a vibronic dimer. (b) Diagrammatic representation of the electronic states ($\vert A \rangle, \vert D \rangle$) and the corresponding spectral densities in the electronic basis. (c) Diagramatic representation of the vibronic basis extended up to the third vibronic sub-block with corresponding spectral densities. \label{fig_basis}}
	\end{center}
\end{figure*}

Figure \ref{fig_basis}a provides a schematic representation of the minimal heterodimer studied here. For a model heterodimer, the electronic excitation of the donor ($E_d$, $\vert D \rangle$) and acceptor ($E_a$, $\vert A \rangle$ ) pigments have an energy gap much larger than the electronic coupling ($E_d - E_a \gg V$). The electronic states of both pigments are coupled to an independent collections of low-frequency vibrations (`electronic environment', Fig. \ref{fig_basis}b/c) that form a thermal bath described by an overdamped Brownian oscillator spectral density
\begin{equation}
J_{\rm elec}(\omega) = 2\lambda_{\rm elec}\frac{\omega\gamma_{\rm elec}}{\omega^2+\gamma^2_{\rm elec}},
\end{equation}
where $\lambda_{\rm elec}$ is the reorganization energy and $\gamma_{\rm elec}$ is the peak width. The electronic state of the acceptor pigment is also coupled to a high-frequency vibration that supports direct vibronic transport with the detuned donor. The high-frequency vibration is, in turn, coupled to a continuum of vibrational modes that form a thermal bath and cause the relaxation of vibrational excitations. In the electronic basis (Fig. \ref{fig_basis}b) the system states are the electronic states of the pigments and the combined influence of the high-frequency vibration and its thermal bath are represented by an effective underdamped Brownian oscillator spectral density (`effective vibration', Fig. \ref{fig_basis}b) 
\begin{equation} \label{eq:J_eff}
J_{\rm eff}(\omega) = 2\lambda_{\rm vib} \frac{2\, \gamma_{\rm vib}\, \Omega_{\rm vib}^2\, \omega}{(\Omega_{\rm vib}^2-\omega^2)^2+4\gamma_{\rm vib}^2\omega^2}, 
\end{equation}
where $\lambda_{\rm vib} = S\cdot \Omega_{\rm vib}$ is the reorganization energy, S is the Huang-Rhys factor, $\gamma_{\rm vib}$ is the peak width, and $\Omega_{\rm vib}$ is the vibrational frequency. 

While simulations performed in the electronic basis can provide an exact description of the net excitation transport between the donor and acceptor they cannot provide clear insight into the underlying vibronic mechanism because the dynamics of the resonant vibration are not explicitly described. The vibronic basis (Fig. \ref{fig_basis}c) illuminates the mechanism of vibronic transport by explicitly incorporating the high-frequency vibration into the system Hamiltonian \cite{holstein1959a,Spano2010a}. The resulting system states are indexed by both the electronic state of the dimer and the nuclear quantum number of the high-frequency harmonic oscillator coupled to the acceptor pigment ($\vert A, \nu_e \rangle$, $\vert D, \nu_g \rangle$). Despite the vibration being coupled only to the acceptor (right hand states, Fig. \ref{fig_basis}c), there is also a ladder of vibrational states when the donor is electronically excited (left hand states, Fig. \ref{fig_basis}c) - corresponding to the vibrational excitation of the acceptor in its ground electronic state. When the effective spectral density ($J_{\rm eff}$, eq. \ref{eq:J_eff}) in the electronic basis is described by an underdamped Brownian oscillator, the thermal bath of vibrational modes coupled to the resonant vibration (`vibrational environment', Fig. \ref{fig_basis}c) is described by an Ohmic spectral density \cite{garg1985a} (Supplementary Information section IC), 
\begin{equation}\label{eq:J_vib}
J_{\rm vib}(\omega)= \frac{\gamma_{\rm vib}}{\Omega_{\rm vib}}\omega e^{-\frac{\omega}{\omega_c}},
\end{equation}
where $\omega_c$ is the cut-off frequency assumed to be much larger then the frequency of the resonant vibration ($\omega_c \gg \Omega_{\rm vib}$). The vibrational environment ($J_{\rm vib}$) drives transport between states that have the same electronic indices but vibrational quantum numbers that differ by $\pm 1$ (green arrows, Fig. \ref{fig_basis}c). We note that in the absence of this additional spectral density in the vibronic basis the high-frequency vibration does not thermalize. 

Assuming the low-frequency vibrations form a Markovian thermal environment, vibronic transport depends on four essential parameters (Supplementary Information section ID): the vibronic coupling ($V_{\textrm{vib}} = V \langle 1_e \vert 0_g \rangle \approx V \sqrt{S}$) between the donor and vibrationally excited acceptor states, the energy gap ($\sigma_{\textrm{vib}}$) between the donor and the vibrationally excited acceptor states, the magnitude of thermal fluctuations in the pigment excitation energies driven by the low frequency vibrational environment ($\lambda_{\textrm{elec}} \rightarrow \textrm{rmsd:} \sqrt{\lambda_{\textrm{elec}} \textrm{k}_{\textrm{b}}\textrm{T}}$), and the rate of vibrational relaxation in the high-frequency mode ($\gamma_{\textrm{vib}}$). In what follows, we track the mechanism of vibronic transport across the space of vibronic parameters expressed in reduced units defined by their ratio to the vibronic coupling: $\Sigma_{\textrm{vib}} = \frac{\sigma_{\textrm{vib}}}{V_\textrm{vib}}$, $\Lambda_{\textrm{elec}}= \frac{\lambda_{\textrm{elec}}}{V_\textrm{vib}}$, $\Gamma_{\textrm{vib}}=\frac{\gamma_{\textrm{vib}}}{V_\textrm{vib}}$. We use hierarchically coupled equations of motion (HEOM)\cite{Tanimura1989a,Tanimura2012a}, as implemented in {\it QMaster} \cite{Kreisbeck2014a}, to simulate excitation transport in the electronic basis and vibronic Redfield \cite{Maly2016a} to simulate transport in the vibronic basis. In supplemental information section IIIB, we demonstrate that vibronic Redfield simulations reproduce HEOM dynamics when the low-frequency vibrations are Markovian, as we assume in all calculations presented here. Computational details are given in Supplemental information section II. We will begin by considering the case where the donor and vibrationally excited acceptor states have the same energy ($\Sigma_{\textrm{vib}} = 0$), and, therefore, vibronic transport is a resonant process. This represents the best-case scenario for coherent vibronic transport. In reality, of course, molecular complexes experience a disordered ensemble of configurations with a distribution of energy gaps between the donor and vibrationally excited acceptor states; we will return at the end to consider how disorder influences optimal behavior in photosynthetic pigment protein complexes and engineered devices. 

\begin{figure}[t!]
	\begin{center}
		\includegraphics{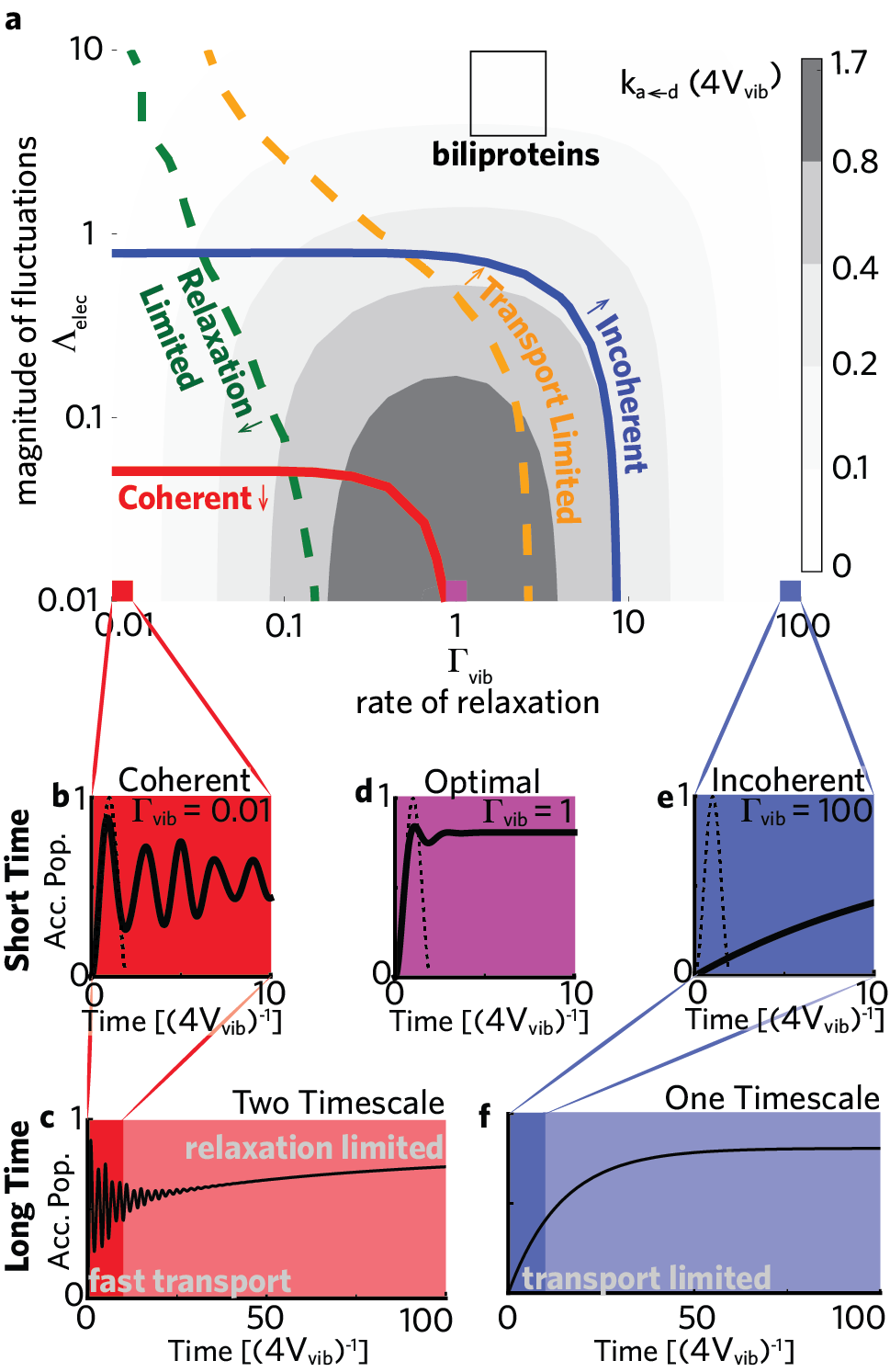}
		\caption{(a) The overall rate from the donor to acceptor is plotted as a function of the magnitude of thermal fluctuations ($\Lambda_{\textrm{elec}}$) and the rate of relaxation ($\Gamma_{\textrm{vib}}$) when the donor and vibrationally excited acceptor states are in resonance ($\Sigma_{\textrm{vib}} =0$.) Mechanistic regimes are bounded by colored lines. The left (right) of the red (blue) solid line is the coherent (incoherent) regime. To the left (right) of the green (orange) dashed-line dynamics are relaxation (transport) limited. The phycobiliprotein parameters estimated experimentally all fall into the black box in the incoherent, transport-limited regime. (b-f) Acceptor populations dynamics simulated with HEOM for different values of coupling to the thermal environments. The dashed line corresponds to one period of unitary dynamics between the donor and vibrationally excited acceptor states. These calculations use $V_{\textrm{vib}} = 0.788$ cm$^{-1}$, $\gamma_{\textrm{elec}} = 50$ cm$^{-1}$, and E$_{d}$ - E$_{a}$ = 350 cm$^{-1}$. \label{figr_dimer}}
	\end{center}
\end{figure}

The defining feature of coherent vibronic transport in a heterodimer is transient delocalization between the donor and the vibrationally excited acceptor states which has been hypothesized to enhance the overall rate of transport between the detuned donor and acceptor pigments (k$_{\rm a \leftarrow d}$) compared to an incoherent vibronic hopping mechanism. Figure \ref{figr_dimer}a shows a contour plot of k$_{\rm a \leftarrow d}$ when the donor and vibrationally excited acceptor state are in resonance ($\Sigma_{\rm vib} = 0$) as a function of the magnitude of thermal fluctuations ($\Lambda_{\textrm{elec}}$) and the rate of relaxation ($\Gamma_{\textrm{vib}}$). Here, we determine k$_{\rm a \leftarrow d}$ by a one-exponential fit to the total acceptor population (summed over all vibrational states) simulated with HEOM (Supplementary Information section IIIA). We find k$_{\rm a \leftarrow d}$ is maximized when the rate of vibrational relaxation is comparable to the vibronic coupling ($\Gamma_{\rm vib} \approx 1$) and there are minimal thermal fluctuations from the low-frequency vibrational modes ($\Lambda_{\rm vib} \approx 0$). The corresponding total acceptor population dynamics show a rapid rise followed by minimal subsequent oscillations (solid line, Fig. \ref{figr_dimer}d). The initial rise in acceptor population dynamics is similar to the dynamics expected between the donor and vibrationally excited acceptor in the absence of any thermal environments (dashed line, Fig. \ref{figr_dimer}d). This similarity suggests that these dynamics correspond to an early time delocalization between the donor and acceptor pigments supported by the resonant vibration. 

\begin{figure}[t!]
	\begin{center}
		\includegraphics{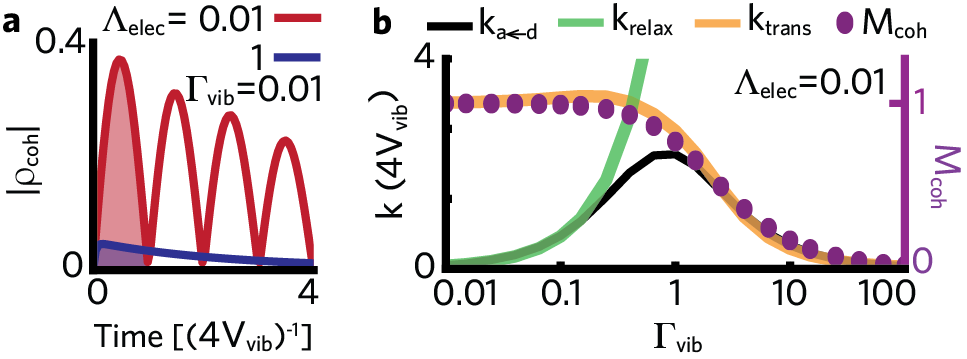}
		\caption{(a) The $\vert \rho_{\textrm{coh}}\vert$ calculated by vibronic Redfield as a function of time. (b) The overall rate from the donor to acceptor is superimposed with the best-fit vibronic transport ($k_{\textrm{trans}}$) and relaxation ($k_{\textrm{relax}}$) rates as a function of $\Gamma_{\textrm{vib}}$ when $\Lambda_{\rm elec} = 0.01$. Other parameters are the same as in Fig. \ref{figr_dimer}.  \label{figd_regimes}}
	\end{center}
\end{figure}

We quantify the extent of early-time delocalization between the donor and vibrationally excited acceptor states by integrating the absolute value of the corresponding off-diagonal element of the vibronic Redfield density matrix ($\rho_{\textrm{coh}} = \vert A, 1_e \rangle \langle D, 0_g \vert$) over one half-period of the acceptor population oscillation ($T_{\textrm{coh}}$), shown as a colored region in Fig. \ref{figd_regimes}a. 
\begin{equation}\label{eq:Mcoh}
M_{\textrm{coh}} = \frac{\int_{0}^{T_{\textrm{coh}}} \vert \rho_{\textrm{coh}}(t) \vert dt}{\int_{0}^{T_{\textrm{coh}}} \vert \rho^{\Lambda = 0, \Gamma = 0}_{\textrm{coh}}(t) \vert dt}
\end{equation}
The extent of coherence decreases as a function of increasing $\Gamma_{\textrm{vib}}$, leading to incoherent transport when $\Gamma_{\textrm{vib}} = 10$ even if coupling to the electronic environment remains weak (purple circles, right axis, Fig.~\ref{figd_regimes}b). The same transition to an incoherent mechanism occurs more rapidly with increasing magnitude of thermal fluctuations driven by the low-frequency vibrations ($\Lambda_{\textrm{elec}}$), where we find incoherent transport begins by $\Lambda_{\textrm{elec}} \approx 1$. The transition from coherent ($M_{\textrm{coh}}>0.8$) to incoherent ($M_{\textrm{coh}}<0.2$) dynamics is reflected in the early time behavior of the acceptor population (Fig. \ref{figr_dimer}b,d,e): for coherent dynamics, the initial donor excitation performs ballistic transport resulting in at least one large sinusoidal oscillation of acceptor population; for incoherent dynamics the donor excitation performs incoherent hopping transport resulting in a slower exponential rise.  

In a vibronic heterodimer the energy gap between the electronic donor and acceptor states is large enough ($E_d - E_a \sim \Omega_{\textrm{vib}} \gg V, \lambda_{\textrm{elec}}$) that transport following an initial donor excitation has only two components: vibronic transport between nearly degenerate donor and acceptor states (e.g. $ \vert D, 0_g \rangle \rightarrow \vert A, 1_e \rangle $) and vibrational relaxation (e.g. $ \vert A, 1_e \rangle \rightarrow \vert A, 0_e \rangle$). We decompose vibronic dynamics into these two fundamental processes by fitting rate matrix parameters to Redfield population curves for vibrational states up to $\nu_{e/g} =2$, as described in Supplementary Information section IIIC. Fig.~\ref{figd_regimes}b shows the best-fit vibronic transport rate (k$_{\textrm{trans}}$, orange lines, left axis), vibrational relaxation rate (k$_{\textrm{relax}}$, green lines, left axis), and the overall donor-to-acceptor rate (k$_{\rm a \leftarrow d}$) when thermal fluctuations are small ($\Lambda_{\textrm{elec}} = 0.01$). When k$_{\textrm{trans}}$ (k$_{\textrm{relax}}$) is the smaller of the two best-fit rates and no more that 0.04 larger than k$_{\rm a \leftarrow d}$, we consider the dynamics to be transport (relaxation) limited. When the relaxation rate is slow compared to the vibronic coupling ($\Gamma_{\textrm{vib}}\ll 1$), vibrational relaxation is the rate limiting step to achieve maximal excitation population on the acceptor (green line, Fig.~\ref{figd_regimes}b). Relaxation limited vibronic dynamics explains the two timescales observed in the acceptor population when $\Gamma_{\textrm{vib}} \ll 1$ (Fig.~\ref{figr_dimer}b,c): excitation transports rapidly between the donor and vibrationally excited acceptor states but reaches the the ground vibrational state of the acceptor ($\vert A, 0_e \rangle$) only after the slower process of vibrational relaxation. In the opposite extreme, when the relaxation rate is very fast ($\Gamma_{\textrm{vib}}\gg 1$), vibronic transport is rate limiting (orange line, Fig.~\ref{figd_regimes}b). In the transport limited regime, the total acceptor population (Fig.~\ref{figr_dimer}e,f) shows a single timescale representative of the rate of vibronic transport from the donor to the acceptor which is followed by rapid vibrational relaxation.

We describe the vibronic transport mechanism by the extent of coherence (red and blue solid lines, Fig.~\ref{figr_dimer}a) and the potential for a rate-limiting step (green and orange dashed lines, Fig.~\ref{figr_dimer}a). Vibronic transport in phycobiliproteins is well approximated by an effective dimer model because of the relatively large distances between most bilins \cite{Blau2017a,OReilly2014a}. Surprisingly, using the parameters extracted from spectroscopic measurements, all three phycobiliproteins previously assigned to show coherent vibronic transport \cite{Dean2016a,Kolli2012a,Womick2011a} (Supplementary Information section 4) fall well into the incoherent regime (black box, Fig.~\ref{figr_dimer}a). This is consistent with a recent reanalysis of a phycobiliprotein, PC645, which revealed, using HEOM calculations, an incoherent vibronic transport mechanism \cite{Blau2017a}. While the mechanistic map presented here assumes a rapidly relaxing (i.e. Markovian) thermal environment, the addition of non-Markovian modes would only allow for coherence on timescales short compared to the relaxation timescale $(1/\gamma_{\textrm{elec}})$. Detailed simulations of bilin motion in PC645 suggest that a large inertial componet of nuclear reorganization occurs on a timescale of less than 20 fs, while vibronic transport in both simulation and experiment is found to occur on timescales of more than 500 fs, consistent with a Markovian approximation \cite{Blau2017a}. These results provide strong evidence that vibronic transport in phycobiliproteins is incoherent. 


\begin{figure}[t!]
	\begin{center}
		\includegraphics{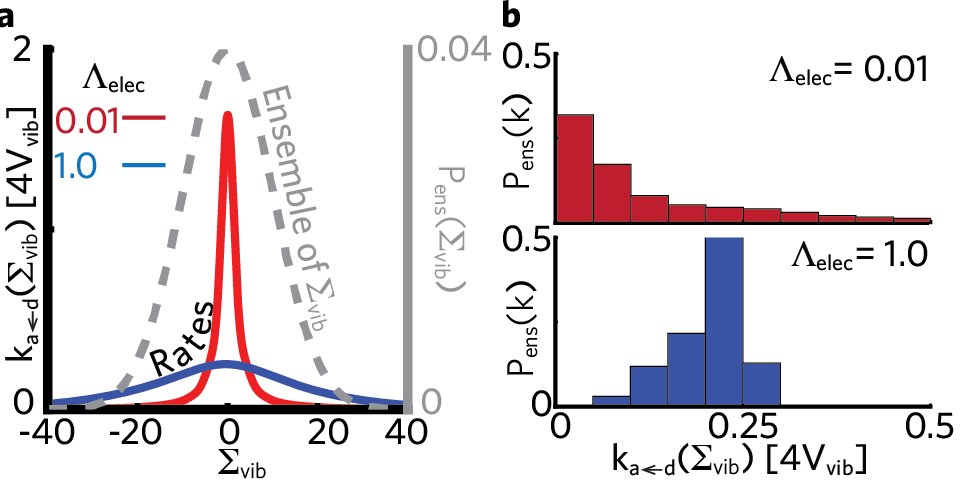}
		\caption{(a) The rate of equilibration (left axis) when $\Gamma_{\textrm{vib}} = 1$ and $\Lambda_{\textrm{elec}} = 0.01$ (red line) or 1 (blue line) is plotted as a function of the energetic detuning between the donor and vibrationally excited acceptor state ($\Sigma_{\textrm{vib}}$). A representative ensemble of $\Sigma_{\textrm{vib}}$ has a standard deviation of 10 (grey dashed line, right axis). (b) The probability of a given rate of equilibration is plotted for $\Lambda_{\textrm{elec}} = 0.01$ (top panel) and $\Lambda_{\textrm{elec}} = 1$ (bottom panel) assuming a Gaussian ensemble of $\Sigma_{\textrm{vib}}$ with standard deviation of 10. Other parameters are the same as in Fig. \ref{figr_dimer}.  \label{figr_disorder}}
	\end{center}
\end{figure}

Does the incoherent nature of vibronic transport in phycobiliproteins point towards inefficiency or an unexplained design principle? Up to this point we have considered the high-frequency vibration to be perfectly resonant with the energy gap between pigments. As a result, we always see the overall donor-to-acceptor rate (k$_{\rm a \leftarrow d}$) decrease with increasing magnitude of thermal fluctuations ($\Lambda_{\textrm{elec}}$). In the presence of an energy gap between the donor and vibrationally excited acceptor ($\Sigma_{\textrm{vib}} \neq 0$), however, increasing the magnitude of thermal fluctuations can increase the fraction of time the two states spend near resonance, thereby enhancing the overall donor-to-acceptor rate (red/blue line , Fig.~\ref{figr_disorder}a) - analogous to the environmentally assisted quantum transport (ENAQT) mechanism observed previously in electronic transport \cite{Ishizaki2009a,Rebentrost2009a}. In practice, molecular aggregates, photosynthetic pigment protein complexes among them, have an ensemble of conformations with slightly different pigment excitation energies leading to energetic disorder. For phycobiliproteins, energetic disorder is expected to have a standard deviations of at least $10$V$_{\textrm{vib}}$ (grey line, Fig.~\ref{figr_disorder}a), as explained in Supplementary Information section IVD. If we calculate the distribution of the overall donor-to-acceptor rate (k$_{\rm a \leftarrow d}$) in a representative disordered ensemble (Supplementary Information section IIID), shown in Fig.~\ref{figr_disorder}b, we find that 50\% of dimers have a rate $<$$0.1$ when $\Lambda_{\textrm{elec}} = 0.01$, compared to only $4\%$ of dimers when $\Lambda_{\textrm{elec}} = 1$. In physiological conditions, absorbed excitation must be successfully transferred through most antenna complexes not merely the set of resonant complexes. In this context, the incoherent vibronic transport mechanism of phycobiliproteins is, in fact, better for functional light harvesting antennae. This points to a basic design principles for vibronic transport: incoherent mechanisms are more robust to disorder and, therefore, are expected to be more efficient when the magnitude of disorder exceeds the vibronic coupling.


The advantage of incoherent vibronic transport in the presence of disorder highlights an important requirement for engineering new materials that use a coherent vibronic mechanism: we must be able to limit the extent of energetic disorder relative to the vibronic coupling. Photosynthetic pigment protein complexes greatly reduce energetic disorder compared to most synthetic aggregates by reproducibly folding with specific pigment positions and orientations, but still we find that at least in some cases disorder can exceed the vibronic coupling. Therefore, we suggest that one of the key challenges for engineering materials to enable coherent vibronic transport is to develop new synthetic structures that dramatically reduce energetic disorder. Finally, our results also suggest that a re-evaluation of the extent of coherence in vibronic transport for other families of photosynthetic pigment protein complexes may be appropriate. The mechanistic map developed here provides a simple rule-of-thumb for determining the extent of coherence in a vibronic heterodimer. Previous work in electronic transport has shown that the transition from coherent to incoherent regimes, used to define the domains of pigments appropriate for generalized F\"{o}rster theory \cite{Raszewski2008a,Novoderezhkin2011a}, follow similar rules-of-thumb to the electronic dimer. In the case of the Fenna-Mathews-Olson (FMO) complex, for example, the Hamiltonian parameters \cite{Nalbach2015a} give $\Lambda_{\rm elec} > 3$ which suggests that an incoherent vibronic mechanism should be expected, consistent with the recent re-interpretation of the 2D optical spectra \cite{Duan02017a}. Thus, we expect our current results provide a foundation for understanding the mechanism of transport in larger photosynthetic aggregates where coherent vibronic transport has been suggested \cite{Nalbach2015a,Romero2014a,Fuller2014a}. 


\begin{acknowledgement}
	We acknowledge the Center for Excitonics, an Energy Frontier Research Center funded by the U.S. Department of Energy, Office of Science and Office of Basic Energy Sciences, under Award Number DE-SC0001088. D.I.G.B., A.A.G., P.M., and R.v.G. acknowledge CIFAR, the Canadian Institute for Advanced Research, for support through the Bio-Inspired Solar Energy program. D.I.G.B. and A.A.G. acknowledge the John Templeton Foundation (Grant Number 60469). We thank Nvidia for support via the Harvard CUDA Center of Excellence. This research used computational time on the Odyssey cluster, supported by the FAS Division of Science, Research Computing Group at Harvard University.
\end{acknowledgement}

\begin{suppinfo}
\textbf{Section I:} Vibronic dimer Hamiltonian, connection between electronic and vibronic basis, \textbf{Section II:} computational details of HEOM simulations, bounds on the high-temperature approximation for underdamped brownian oscillators, computational details of vibronic Redfield simulations, \textbf{Section III:} description of extracting total donor-to-acceptor transport rate ($k_{a \leftarrow d}$), comparison of HEOM and vibronic Redfield simulations in Markovian and less Markovian regimes, description of determining the effective transport and relaxation rates, description of treating disorder ensembles of vibronic dimers, and, finally, \textbf{Section IV:} parameter extraction for effective dimer models of photosynthetic pigment protein complexes. 
\end{suppinfo}

\bibliography{VibDimer.bib}


\section*{Supplementary material}

# Supplementary Information: Mechanistic regimes of vibronic transport in a heterodimer and the design principle of incoherent vibronic transport in phycobiliproteins

Doran I. G. Bennett,[1, 2, *] Pavel Malý,[3, 4, 2] Christoph Kreisbeck,[1] Rienk van Grondelle,[3, 2] and Alan Aspuru-Guzik[1, 2]

[1]*Department of Chemistry and Chemical Biology, Harvard University,*
*12 Oxford St, Cambridge, Massachusetts 02138, USA*
[2]*Bio-Inspired Solar Energy Program, Canadian Institute for Advanced Research, Toronto, Ontario M5G 1Z8, Canada*
[3]*Department of Physics and Astronomy, Faculty of Sciences,*
*VU University Amsterdam, De Boelelaan 1081, 1081 HV Amsterdam, The Netherlands*
[4]*Institute of Physics, Faculty of Mathematics and Physics,*
*Charles University, Ke Karlovu 3, 121 16 Prague 2, Czech Republic*

## Contents

## I. HAMILTONIAN FOR EXCITON TRANSPORT

In the following we describe the spin-boson Hamiltonian solved in the main text. In its general form it is composed of four components (shown schematically in Fig. S1):

- $\mathcal{H}_{\text{elec}}$ describes the electronic states of the pigments (in this case two pigments: donor and acceptor with an electronic coupling V between their excited states),

---

*Electronic address: doranbennett@g.harvard.edu

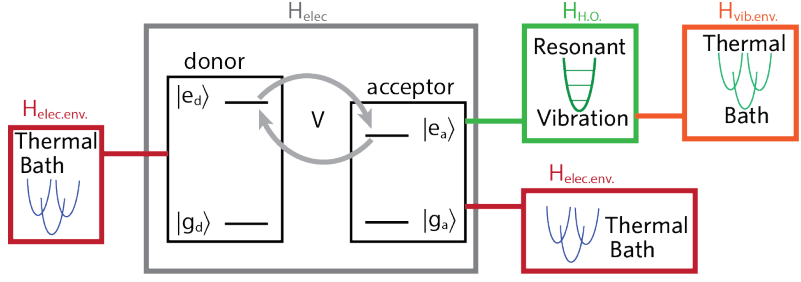

Figure S1: **Schematic Representation of Spin-Boson Hamiltonian for a Vibronic Dimer.**

- $\mathcal{H}_{\text{H.O.}}$ describes a high-frequency vibration and its coupling to the acceptor pigment,

- $\mathcal{H}_{\text{vib.-env.}}$ describes a continuum of vibrational modes and their coupling to the high-frequency vibration described by $\mathcal{H}_{\text{H.O.}}$,

- and, $\mathcal{H}_{\text{elec.-env.}}$ describes the continuum of low-frequency vibrational modes and their coupling to the pigment electronic states (donor and acceptor have independent and identical collections of low-frequency vibrations).

All together, the total Hamiltonian is given by

$$\mathcal{H} = \mathcal{H}_{\text{elec}} + \mathcal{H}_{\text{H.O.}} + \mathcal{H}_{\text{vib.-env.}} + \mathcal{H}_{\text{elec.-env.}}. \tag{S1}$$

Note that both $\mathcal{H}_{\text{vib.-env.}}$ and $\mathcal{H}_{\text{elec.-env.}}$ represent thermal environments composed of a continuum of vibrational modes. In one case, $\mathcal{H}_{\text{vib.-env.}}$, these modes couple to the high-frequency vibration and induce vibrational relaxation. In the other case, $\mathcal{H}_{\text{elec.-env.}}$, these modes couple to the excitation energies of the pigments and induce fluctuations and a relaxation in the excitation energies.

There are typically two ways of applying an open quantum systems approach to solving exciton dynamics described by eq. S1, denoted in the main text as the vibronic and electronic basis. In the electronic basis (see section I A), the system Hamiltonian is composed of only the electronic component ($\mathcal{H}_{\text{sys}} = \mathcal{H}_{\text{elec}}$) and the harmonic nature of the remaining degrees of freedom is used to construct an effective electronic bath model which includes the high-frequency harmonic oscillator ($\mathcal{H}_{\text{H.O.}}$) implicitly. In the vibronic basis (see section I B), we include the high-frequency harmonic oscillator ($\mathcal{H}_{\text{H.O.}}$) explicitly in the system Hamiltonian ($\mathcal{H}_{\text{sys}} = \mathcal{H}_{\text{elec}} + \mathcal{H}_{\text{H.O.}}$), whereas we trace out the degrees of freedoms of the two environments ($\mathcal{H}_{\text{vib.-env.}}$ and $\mathcal{H}_{\text{elec.-env.}}$) to construct the effective bath models. We discuss in detail the relationship between the two basis sets in Section I C.

## A. Hamiltonian expanded in electronic basis

In the electronic basis, the system Hamiltonian is given by just the electronic degrees of freedom,

$$\mathcal{H}_{\text{elec}} = \mathcal{E}_d |D\rangle\langle D| + \mathcal{E}_a |A\rangle\langle A| + V(|D\rangle\langle A| + |A\rangle\langle D|), \tag{S2}$$

where $|D\rangle$ ($|A\rangle$) denotes the excited electronic state of the donor (acceptor) pigment and $\mathcal{E}_d$ ($\mathcal{E}_a$) is the energy gap between the minimum energy geometries of the donor (acceptor) in the ground vs excited electronic state. We use an open quantum systems approach to recast the coupling between the electronic and vibrational degrees of freedom into two kinds of effective vibrational environments. The first kind of vibrational environment arises from the direct coupling of a large number of low-frequency vibrations to the electronic state of the pigment,

$$\mathcal{H}_{\text{elec.-env.}} = \sum_{m \in \{D,A\}} |m\rangle\langle m| \left(\lambda_{\text{elec}} + \sum_i \omega_{i,m} d_{i,m} x_{i,m}\right) + \sum_{m \in \{D,A\}} \frac{1}{2}\omega_{i,m}(P_{x_{i,m}}^2 + x_{i,m}^2), \tag{S3}$$

where $d_{i,m}$ represents the displacement (in reduced units) of the equilibrium position along the coordinate $x_{i,m}$ when the pigment ($m$) is in its ground versus first excited state. The corresponding spectral density is defined by $J_{m \in \{A,D\}}(\omega) = \frac{\pi}{2} \sum_i \omega_{i,m}^2 d_{i,m}^2 \delta(\omega - \omega_{i,m})$. In the following, we assume that the electronic environment can be described by an over-damped Brownian oscillator spectral density,

$$J_{\text{elec}}(\omega) = 2\lambda_{\text{elec}} \frac{\omega \gamma_{\text{elec}}}{\omega^2 + \gamma_{\text{elec}}^2}, \tag{S4}$$

where $\lambda_{\mathrm{elec}}$ is the reorganization energy and $\gamma_{\mathrm{elec}}$ defines the rate at which the vibrational environment responds to a change in the electronic state of the pigment.

The second kind of vibrational environment arises from the coupling of the electronic state of the acceptor pigment to a single high-frequency harmonic oscillator that is, in turn, coupled to a continuum of other vibrational modes,

$$\mathcal{H}_{\mathrm{H.O.}} + \mathcal{H}_{\mathrm{vib.-env.}} = (\lambda_{\mathrm{vib}} + \Omega_{\mathrm{vib}}\sqrt{S}Q)|A\rangle\langle A| + \frac{1}{2}\Omega_{\mathrm{vib}}(P_Q^2 + Q^2) + \sum_k \frac{1}{2}\omega_k p_k^2 + \sum_k \omega_k(q_k + c_k Q)^2, \qquad \text{(S5)}$$

where $\lambda_{\mathrm{vib}}$ is the reorganization energy of the high-frequency mode, $\Omega_{\mathrm{vib}}$ is the central frequency of the vibration, S is the Huang-Rhys factor ($S = \frac{1}{2}\Delta Q^2$), and $\Delta Q$ is the unitless displacement of the minimum energy geometry for the high-frequency vibration between the ground and excited electronic states. We note that the presence of $\frac{1}{2}\omega_k(q_k + c_k Q)^2$ term in the Hamiltonian assumes that a matching $-\sum_k (c_k Q)^2$ term has been incorporated into the definition of the vibrational frequency $\Omega_{\mathrm{vib}}$. In other words, the apparent central frequency of the explicit Harmonic oscillator ($\Omega_{\mathrm{vib}}$) is shifted by the coupling to a thermal environment. The corresponding spectral density is solved by diagonalizing the vibrational states ($\{Q, q_k\}$) and then rewriting eq. S5 in terms of the normal modes ($\tilde{q}_k$),

$$\mathcal{H}_{\mathrm{H.O.}} + \mathcal{H}_{\mathrm{vib.-env.}} = \sum_k \frac{1}{2}\hbar\tilde{\omega}_k \tilde{p}_k^2 + \sum_k \frac{1}{2}\hbar\tilde{\omega}_k \left(\tilde{q}_k + \Delta\tilde{c}_k^2|A\rangle\langle A|\right)^2. \qquad \text{(S6)}$$

The resulting spectral density is $J_{\mathrm{vib}}(\omega) = \frac{\pi}{2}\sum_k \tilde{\omega}_k^2 \tilde{c}_k^2 \delta(\omega - \tilde{\omega}_k)$. In the following, we will assume that the influence of a high-frequency vibration is described by an under-damped Brownian oscillator spectral density,

$$J_{\mathrm{vib}}(\omega) = 2\lambda_{\mathrm{vib}} \frac{2\gamma_{\mathrm{vib}}\Omega_{\mathrm{vib}}^2 \omega}{(\Omega_{\mathrm{vib}}^2 - \omega^2)^2 + 4\gamma_{\mathrm{vib}}^2\omega^2}, \qquad \text{(S7)}$$

where $\gamma_{\mathrm{vib}}$ is the rate of vibrational relaxation in the high-frequency mode driven by coupling to its thermal environment ($\mathcal{H}_{\mathrm{vib.-env.}}$).

## B. Hamiltonian expanded in vibronic basis

In the vibronic basis, the system Hamiltonian is composed of both the electronic degrees of freedom and the high-frequency vibrational mode,

$$\begin{aligned}\mathcal{H}_{\mathrm{sys}} = \mathcal{H}_{\mathrm{elec}} + \mathcal{H}_{\mathrm{H.O.}} =& \mathcal{E}_d|D\rangle\langle D| + \mathcal{E}_a|A\rangle\langle A| + V(|D\rangle\langle A| + |A\rangle\langle D|) \\ &+ (\Omega_{\mathrm{vib}}\sqrt{S}Q + \lambda_{\mathrm{vib}})|A\rangle\langle A| + \frac{1}{2}\Omega_{\mathrm{vib}}(P_Q^2 + Q^2)\end{aligned} \qquad \text{(S8)}$$

where the parameters have the same definitions as for eqs. S2 and S5. The resulting system states are described by both the electronic indices of the pigments and the vibrational quantum number of the high-frequency mode ($|D, \nu_g\rangle$, $|A, \nu_e\rangle$). The sub-index on the vibrational quantum number ($\nu_g$, $\nu_e$) specifies if the vibrational mode is on the ground or excited state potential energy surface of the acceptor pigment.

There are, again, two spectral density components that describe the influence of vibrational modes not included in the system Hamiltonian. The first component represents the same low-frequency vibrational modes that directly couple to the pigment electronic states that were described in the electronic basis,

$$\mathcal{H}_{\mathrm{elec.-env.}} = \sum_{m \in \{D, A\}} |m\rangle\langle m| \left(\lambda_{\mathrm{elec}} + \sum_i \omega_{i,m} d_{i,m} x_{i,m}\right) + \sum_{m \in \{D, A\}, i} \frac{1}{2}\omega_{i,m}(P_{\alpha_{i,m}}^2 + x_{i,m}^2), \qquad \text{(S9)}$$

which we again describe by an over-damped Brownian oscillator spectral density,

$$J_{\mathrm{elec}}(\omega) = 2\lambda_{\mathrm{elec}} \frac{\omega\gamma_{\mathrm{elec}}}{\omega^2 + \gamma_{\mathrm{elec}}^2}. \qquad \text{(S10)}$$

This spectral density results in fluctuations of the pigment excitation energies.

The second spectral density describes the influence of the continuum of vibrational modes that couple to the high-frequency vibration,

$$\mathcal{H}_{\mathrm{vib.-env.}} = \sum_k \frac{1}{2}\omega_k p_k^2 + \sum_k \frac{1}{2}\omega_k(q_k + c_k Q)^2. \qquad \text{(S11)}$$

In this case, the coupling to the system Hamiltonian arises from the $(q_k + c_k Q)^2$ term and results in a fluctuating coupling between states with the same electronic indices but different vibrational quantum numbers. In Section I C we demonstrate, as has been previously shown [1], that an underdamped Brownian oscillator spectral density in the electronic basis is reproduced by an Ohmic spectral density in the vibronic basis

$$J_{\text{vib}} = \frac{\pi}{2} \sum_k \omega_k^2 c_k^2 \delta(\omega - \omega_k) = \lim_{\omega_c \to \infty} \frac{\gamma_{\text{vib}}}{\Omega_{\text{vib}}} \omega \exp(-\omega/\omega_c). \tag{S12}$$

We note that in the absence of this second spectral density there would be no vibrational relaxation in the high-frequency harmonic oscillator, so this term is essential for correctly reproducing vibronic dynamics.

## C. Mapping between $J_{\text{eff}}(\omega)$ and $J_{\text{vib}}(\omega)$

The equivalence of the effective spectral density in the electronic basis and an explicit harmonic oscillator coupled to an ohmic spectral density in the vibronic basis is well-established in literature [1–4]. Notice that the difference between the spectral density in the electronic basis versus the vibronic basis is the expansion of $\{Q, q_k\}$ into normal modes $\{\tilde{q}_k\}$ and does not depend on the exciton degrees of freedom. The defining feature of the spectral density representing the high-frequency mode in the electronic basis is that it is an effective representation of the combined influence of $\mathcal{H}_{\text{H.O.}} + \mathcal{H}_{\text{vib.-env.}}$. We follow Garg, et al. [1] and Leggett, et al. [2] in deriving the effective spectral density in the electronic basis ($J_{\text{eff}}$) from $J_{\text{vib}}$ by solving the equations of motion for a general particle with position coordinate $\alpha$ bound in a arbitrary potential $U(\alpha)$ with bi-linear coupling to the high-frequency vibrational mode Q. The resulting Hamiltonian is

$$\mathcal{H}_\alpha = \frac{p_\alpha^2}{2\mu} + U(\alpha) + \frac{1}{2}\Omega_{\text{vib}}\big(P^2 + (Q + \Delta Q\alpha)^2\big) + \sum_k \frac{1}{2}\omega_k\big(p_k^2 + (q_k + c_k Q)^2\big). \tag{S13}$$

As is shown by Leggett [2] the effective spectral density $J_{\text{eff}}(\omega)$ can be extracted by analyzing the classical equations of motion

$$\mu\ddot{\alpha} = -\frac{dU(\alpha)}{d\alpha} - \Delta Q\Omega_{\text{vib}}(Q + \Delta Q\alpha) \tag{S14}$$

$$\frac{1}{\Omega_{\text{vib}}}\ddot{Q} = -\Omega_{\text{vib}}(Q + \Delta Q\alpha) - \sum_k c_k\omega_k q_k - Q\sum_k c_k^2\omega_k \tag{S15}$$

$$\frac{1}{\omega_k}\ddot{q}_k = -\omega_k q_k - Q\omega_k c_k. \tag{S16}$$

Applying the Fourier transformation to the above equations results in a set of coupled algebraic equations in frequency domain

$$\big(-\mu\omega^2 + \Delta Q^2\Omega_{\text{vib}}\big)\alpha(\omega) + \Delta Q\Omega_{\text{vib}}Q(\omega) = -U'_\omega(\alpha) \tag{S17}$$

$$\Big(-\frac{1}{\Omega_{\text{vib}}}\omega^2 + \Omega_{\text{vib}} + \sum_k c_k^2\omega_k\Big)Q(\omega) + \sum_k c_k\omega_k q_k(\omega) = -\Omega_{\text{vib}}\Delta Q\alpha(\omega) \tag{S18}$$

$$q_k(\omega) = -\frac{\omega_k^2 c_k}{\omega_k^2 - \omega^2}Q(\omega), \tag{S19}$$

where $\alpha(\omega)$, $Q(\omega)$, $q_k(\omega)$ and $U'_\omega(\alpha)$ are the Fourier transforms of $\alpha(t)$, $Q(t)$, $q_k(t)$ and $\frac{dU(\alpha)}{d\alpha}$.

The effective spectral density is defined by the equation of motion for the general particle position in the frequency domain ($\alpha(\omega)$),

$$K(\omega)\alpha(\omega) = -U'_\omega(\alpha) \tag{S20}$$

$$J_{\text{eff}}(\omega) = \lim_{\epsilon \to 0+}\text{Im}K(\omega - i\epsilon). \tag{S21}$$

we derive $K(\omega)$ by combining eq. (S19) with eq. (S18)

$$\Big(-\frac{1}{\Omega_{\text{vib}}}\omega^2 + \Omega_{\text{vib}} - \omega^2\sum_k c_k^2\omega_k\frac{1}{\omega_k^2 - \omega^2}\Big)Q(\omega) = -\Omega_{\text{vib}}\Delta Q\alpha(\omega), \tag{S22}$$

and solving for $Q(\omega)$

$$Q(\omega) = \frac{-\Omega_{\mathrm{vib}}\Delta Q}{\Omega_{\mathrm{vib}} + L(\omega)}\alpha(\omega) \tag{S23}$$

where we have defined

$$L(\omega) = -\omega^2\Big(\frac{1}{\Omega_{\mathrm{vib}}} + \sum_k c_k^2 \omega_k \frac{1}{\omega_k^2 - \omega^2}\Big). \tag{S24}$$

Combining eq. S23 with eq. S17 we find that

$$\big(-\mu\omega^2 + \Delta Q^2\Omega_{\mathrm{vib}}\big)\alpha(\omega) - \frac{\Omega_{\mathrm{vib}}^2\Delta Q^2}{\Omega_{\mathrm{vib}} + L(\omega)}\alpha(\omega) = -U_\omega'(\alpha) \tag{S25}$$

which can be written into the form of eq. S20 to yield

$$K(\omega) = -\mu\omega^2 + \frac{\Delta Q^2\Omega_{\mathrm{vib}}L(\omega)}{\Omega_{\mathrm{vib}} + L(\omega)}. \tag{S26}$$

In order to express $K(\omega)$, which is directly related to the effective spectral density in the electronic basis, we need to solve for $L(\omega)$ in terms of the spectral density in the vibronic basis ($J_{\mathrm{vib}}$). Using the definition of the spectral density $J_{\mathrm{vib}}(\omega) = \frac{\pi}{2}\sum_k c_k^2 \omega_k^2 \delta(\omega - \omega_k)$, taking the continuum limit of the summation over the vibrational modes, and assuming an ohmic spectral density with exponential cut-off

$$J_{\mathrm{vib}}(\omega) = \frac{\gamma_{\mathrm{vib}}}{\Omega_{\mathrm{vib}}}\omega\exp(-\omega/\omega_c), \tag{S27}$$

leads to

$$\begin{aligned}
L(\omega) &= -\omega^2\Big(\frac{1}{\Omega_{\mathrm{vib}}} + \frac{2}{\pi}\int d\omega' J(\omega')\frac{1}{\omega'(\omega'^2 - \omega^2)}\Big) \\
&= -\omega^2\Big(\frac{1}{\Omega_{\mathrm{vib}}} + \frac{\gamma_{\mathrm{vib}}}{\Omega_{\mathrm{vib}}}\frac{2}{\pi}\int d\omega'\frac{1}{\omega'^2 - \omega^2}\exp(-\omega/\omega_c)\Big).
\end{aligned} \tag{S28}$$

We carry out the integration employing the Cauchy's residue theorem and obtain a closed expression

$$L(\omega) = -\frac{\omega^2}{\Omega_{\mathrm{vib}}} + i\frac{2\gamma_{\mathrm{vib}}}{\Omega_{\mathrm{vib}}}\omega\exp(-\omega/\omega_c), \tag{S29}$$

which reduces in the Ohmic limit $\omega_c \to \infty$ to

$$L(\omega) = -\frac{\omega^2}{\Omega_{\mathrm{vib}}} + i\frac{2\gamma_{\mathrm{vib}}}{\Omega_{\mathrm{vib}}}\omega. \tag{S30}$$

This leads to the well known Brownian oscillator spectral density in the electronic basis

$$J_{\mathrm{eff}}(\omega) = \Delta Q^2\Omega_{\mathrm{vib}}\frac{2\gamma_{\mathrm{vib}}\Omega_{\mathrm{vib}}^2\omega}{(\Omega_{\mathrm{vib}}^2 - \omega^2)^2 + 4\gamma_{\mathrm{vib}}^2\omega^2} \tag{S31}$$

which can be further simplified by noting that $\Delta Q^2 = 2S$ and $\lambda_{\mathrm{vib}} = S\Omega_{\mathrm{vib}}$

$$J_{\mathrm{eff}}(\omega) = 2\lambda_{\mathrm{vib}}\frac{2\gamma_{\mathrm{vib}}\Omega_{\mathrm{vib}}^2\omega}{(\Omega_{\mathrm{vib}}^2 - \omega^2)^2 + 4\gamma_{\mathrm{vib}}^2\omega^2}. \tag{S32}$$

### D. Building a picture of transport for the vibronic heterodimer

While the electronic and vibronic basis describe the same molecular Hamiltonian (eq. S1), the appropriate physical picture for transport described in the vibronic basis can appear more complicated. In Fig. 1c of the main text, we present a diagramatic representation of vibronic transport in a heterodimer. In the following we outline how this

picture of vibronic transport arises from the the molecular Hamiltonian and connect the pertinent parameters with their counterparts in the electronic basis.

The system Hamiltonian in the vibronic basis is described completely by

$$
\begin{aligned}
\mathcal{H}_{\text{sys}} = \mathcal{H}_{\text{elec}} + \mathcal{H}_{\text{H.O.}} = & \mathcal{E}_d |D\rangle\langle D| + \mathcal{E}_a |A\rangle\langle A| + V(|D\rangle\langle A| + |A\rangle\langle D|) \\
& + (\Omega_{\text{vib}}\sqrt{S}Q + \lambda_{\text{vib}})|A\rangle\langle A| + \frac{1}{2}\Omega_{\text{vib}}(P_Q^2 + Q^2),
\end{aligned}
\tag{S33}
$$

but can also be described in an explicit basis of vibronic states ($|D, \nu_g\rangle, |A, \nu_e\rangle$). Here, we expanded the Hamiltonian up to $\nu_e = 2, \nu_g = 1$ giving

$$
H_{\text{sys}} = \begin{array}{c|ccccc}
 & |A, 0_e\rangle & |D, 0_g\rangle & |A, 1_e\rangle & |D, 1_g\rangle & |A, 2_e\rangle \\
\hline
 & E_a - \lambda_{\text{vib}} & V\langle 0_e|0_g\rangle & 0 & V\langle 0_e|1_g\rangle & 0 \\
 & V\langle 0_e|0_g\rangle & E_d & V\langle 1_e|0_g\rangle & 0 & V\langle 2_e|0_g\rangle \\
 & 0 & V\langle 1_e|0_g\rangle & E_a - \lambda_{\text{vib}} + \Omega_{\text{vib}} & V\langle 1_e|1_g\rangle & 0 \\
 & V\langle 0_e|1_g\rangle & 0 & V\langle 1_e|1_g\rangle & E_d, + \Omega_{\text{vib}} & V\langle 2_e|1_g\rangle \\
 & 0 & V\langle 2_e|0_g\rangle & 0 & V\langle 2_e|1_g\rangle & E_a - \lambda_{\text{vib}} + 2\Omega_{\text{vib}}
\end{array}
$$

where $E_d$ is the vertical excitation energy of the donor, $E_a$ is the vertical excitation energy of the acceptor, and $\langle \nu_e | \mu_g \rangle$ are vibrational overlap integrals (Frank-Condon factors). While further states could be included in the expansion, in our calculations $\beta\Omega_{\text{vib}} > 1$ ensures that the excitation transport dominantly occurs in just the three lowest energy states ($|A, 0_e\rangle, |D, 0_g\rangle, |A, 1_e\rangle$). The vibronic Hamiltonian for a heterodimer has an approximately block-diagonal form with pairs of nearly degenerate donor and acceptor states occurring when $\nu_g = \nu_e - 1$.

There are three possible kinds of transport processes in a vibronic heterodimer: electronic transport between states with different electronic indices but the same vibrational quantum number (e.g. $|D, 0_g\rangle \to |A, 0_e\rangle$), vibronic transport between states with different electronic indices and different vibrational quantum numbers (e.g. $|D, 0_g\rangle \to |A, 1_e\rangle$), and vibrational relaxation between states with the same electronic indices but different vibrational quantum numbers (e.g. $|A, 1_e\rangle \to |A, 0_e\rangle$). Vibronic transport moves excitation between the donor and acceptor within a vibronic sub-block formed by nearly degenerate donor and acceptor states (e.g. $|D, 0_g\rangle, |A, 1_e\rangle$). The only processes by which excitation can move between sub-blocks are electronic transport and vibrational relaxation. Since the coupling between different sub-blocks of the vibronic Hamiltonian is much smaller than the energy gap ($\frac{V}{\Delta E} \ll 1$), electronic transport between the donor and acceptor can only occur if thermal fluctuations driven by the electronic environment are comparable in magnitude to the energy gap between the sub-blocks ($\approx \Omega_{\text{vib}}$). In order to ensure that we are exploring vibronic transport mechanisms and not direct electronic transport across an energy gap, for all calculations presented here we have checked that in the absence of the high-frequency mode there is essentially no transport between the donor and acceptor. Vibrational relaxation on the other hand arises from the interaction between the high-frequency vibration and its thermal environment

$$
\mathcal{H}_{\text{vib.-env.}} = \sum_k \frac{1}{2}\omega_k p_k^2 + \sum_k \frac{1}{2}\omega_k(q_k + c_k Q)^2 = \sum_k \frac{1}{2}\omega_k p_k^2 + \sum_k \frac{1}{2}\omega_k(q_k^2 + 2c_k q_k Q + c_k^2 Q^2)
\tag{S34}
$$

which does not depend on the coupling present in the system Hamiltonian. The bi-linear term ($\sum_k \omega_k c_k q_k Q$) directly drives vibrational relaxation by coupling states with the same electronic index but vibrational quantum numbers that differ by $\pm 1$.

Since neither vibronic transport nor vibrational relaxation depends on the non-block diagonal terms in the system Hamiltonian, there are only three kinds of parameters in the system Hamiltonian that are relevant for the vibronic transport mechanism:

1. $\sigma_n = E_d - E_a + \lambda_{\text{vib}} - \Omega_{\text{vib}}$: the energy gap between between states within a vibronic sub-block,

2. $V_{\text{vib}}^{(n)} = V\langle n_e|(n-1)_g\rangle$: The vibronic coupling between the nearly degenerate donor and acceptor states in the $n^{th}$ sub-block,

3. $\Delta E_n = \Omega_{\text{vib}}$ or $\frac{1}{2}(E_d - E_a + \lambda_{\text{vib}} + \Omega_{\text{vib}})$: The energy gap between the average energies of the vibronic sub-blocks.

In the main text we do not vary the values of $\Delta E$ as this changes the thermodynamic driving force of vibrational relaxation but does not change the mechanism. On the other hand, variations in $\sigma_n$ and $V_{\text{vib}}^{(n)}$ can dramatically change the overall vibronic transport dynamics. We note that the vibronic coupling changes between the different sub-blocks of the system Hamiltonian. In the main text, we have defined $V_{\text{vib}} = V_{\text{vib}}^{(1)}$ which is a useful approximation as long as the majority of excitation is transported between the donor and acceptor within the first vibronic sub-block, as

should be expected in our calculations since $\beta\Omega_{\mathrm{vib}} > 1$. As we shall see when discussing the best-fit vibronic transport rate in section III C below, however, our calculations still show some influence from the larger vibronic coupling in the higher lying sub-blocks.

We note that while the non-block-diagonal elements of the system Hamiltonian in the vibronic basis do not directly influence transport they can result in slight changes in the resonance conditions by shifting the energy levels of the Hamiltonian. When needed, this can be accounted for perturbatively since the couplings remain much smaller than the energy gaps. At second order in perturbation theory the vibronic system Hamiltonian can be expressed as

$$
\tilde{H}_{\mathrm{sys}} =
\begin{array}{c}
\begin{array}{ccccc}
|A,0_e\rangle & |D,0_g\rangle & |A,1_e\rangle & |D,1_g\rangle & |A,2_e\rangle
\end{array} \\
\begin{bmatrix}
\tilde{E}_{a,0} & 0 & 0 & 0 & 0 \\
0 & \tilde{E}_{d,0} & V\langle 1_e|0_g\rangle & 0 & 0 \\
0 & V\langle 1_e|0_g\rangle & \tilde{E}_{a,1}+\Omega_{\mathrm{vib}} & 0 & 0 \\
0 & 0 & 0 & \tilde{E}_{d,1}+\Omega_{\mathrm{vib}} & V\langle 2_e|1_g\rangle \\
0 & 0 & 0 & V\langle 2_e|1_g\rangle & \tilde{E}_{a,2}+2\Omega_{\mathrm{vib}}
\end{bmatrix}
\end{array}
$$

where $\tilde{E}_{a,\nu_e}$ and $\tilde{E}_{d,\nu_g}$ are the perturbatively corrected state energies given by

$$
\tilde{E}_{a,\nu_e} = E_a - \lambda_{\mathrm{vib}} + \nu_e\Omega_{\mathrm{vib}} + \sum_{\mu_g=0,\mu_g\neq\nu_e-1}^{\infty} \frac{|V\langle\nu_e|\mu_g\rangle|^2}{(E_a-\lambda_{\mathrm{vib}}+\nu_e\Omega_{\mathrm{vib}})-(E_d+\mu_g\hbar\Omega_{\mathrm{vib}})}, \tag{S35}
$$

$$
\tilde{E}_{d,\nu_g} = E_d + \nu_g\Omega_{\mathrm{vib}} + \sum_{\mu_e=0,\mu_e\neq\nu_g+1}^{\infty} \frac{|V\langle\mu_e|\nu_g\rangle|^2}{(E_d+\nu_g\Omega_{\mathrm{vib}})-(E_a-\lambda_{\mathrm{vib}}+\mu_e\hbar\Omega_{\mathrm{vib}})}. \tag{S36}
$$

For all calculations presented here the perturbative correction is negligible as a result of the very small value of $\frac{V}{\Delta E} = 0.01$.

## II. SIMULATING EXCITON TRANSPORT

We use both the hierarchically coupled equations of motion (HEOM) approach and vibronic Redfield theory to simulate exciton dynamics. Below we begin by outlining the equations of motion for HEOM and the parameters used in the present simulations. We demonstrate the appropriateness of using shifted Drude-Lorentz peaks in place of underdamped Brownian oscillator spectral densities and then describe the bounds on the high-temperature approximation used in the HEOM simulations. Finally we describe the vibronic Redfield equations of motion and outline the parameters we used for those calculations.

### A. HEOM: Computational details

We simulate the exciton dynamics in the electronic basis with the hierarchically coupled equations of motion (HEOM) approach [5, 6]. HEOM is a numerically exact open quantum system approach, which takes into account the non-perturbative and non-Markovian nature of the exciton phonon coupling. While HEOM in its simplest version is restricted to a Drude-Lorentz spectral density, the method can be easily generalized [7, 8] by decomposing structured spectral densities in form of shifted Drude-Lorentz peaks

$$
J(\omega) = \sum_{k=1}^{M} \left[ \frac{\gamma_k\lambda_k\omega}{\gamma_k^2+(\omega+\Omega_k)^2} + \frac{\gamma_k\lambda_k\omega}{\gamma_k^2+(\omega-\Omega_k)^2} \right]. \tag{S37}
$$

The derivation of HEOM starts from the Liouville-von Neumann equation

$$
\frac{\mathrm{d}}{\mathrm{d}t}R(t) = -\frac{\mathrm{i}}{\hbar}[\mathcal{H}(t),R(t)] = -\frac{\mathrm{i}}{\hbar}\mathcal{L}(t)R(t) \tag{S38}
$$

for the total density matrix $R(t)$, describing both electronic and vibrational degrees of freedom. We assume that the system was initially prepared such that $R(t_0) = \rho(t_0)\otimes\rho_{\mathrm{phon}}(t_0)$ factorizes in electronic an vibrational states. Tracing out the vibrational degrees of freedom results in a formal solution for the time evolution of the reduced density matrix

$$
\rho(t) = \langle \mathrm{T}_+ \exp\left(-\frac{\mathrm{i}}{\hbar}\int_0^t \mathrm{d}s\, \mathcal{L}(s)\right)\rangle \rho(0). \tag{S39}
$$

By employing the Gaussian nature of the phonon bath as well as the exponential form of the two-time correlation function, the time non-local eq. (S39) can be cast into a set of coupled time local equations of motion

$$
\begin{aligned}
\frac{\mathrm{d}}{\mathrm{d}t}\sigma^{(n_1,\ldots,n_N)}(t) \;=\; & \Big(-\frac{\mathrm{i}}{\hbar}\mathcal{L}_{\mathrm{ex}} - \sum_m n_m\gamma\Big)\sigma^{(n_1,\ldots,n_N)}(t) \\
& + \sum_m \frac{\mathrm{i}}{\hbar}V_m^{\times}\sigma^{(n_1,\ldots,n_m+1,\ldots,n_N)}(t) \\
& + \sum_m n_m\theta_m\sigma^{(n_1,\ldots,n_m-1,\ldots,n_N)}(t).
\end{aligned}
\tag{S40}
$$

We also impose a high-temperature approximation by neglecting contributions to the correlation function arising from the Matsubara frequencies. We discuss the appropriateness of this approximation for the current calculations in section II C. We define $\rho(t) = \sigma^{(0,\ldots,0)}(t)$, $\theta_m = \mathrm{i}\Big(\frac{2\lambda}{k_B T\hbar}V_m^{\times}(t) - \mathrm{i}\lambda\gamma V_m^{\circ}(t)\Big)$, $V_m^{\times}\bullet = [V_m,\bullet]$, $V_m^{\circ}\bullet = \{V_m,\bullet\}$ and $V_m = |m\rangle\langle m|$. For the model dimer discussed in the main text the operators $V_m$ read $V_1 = |1\rangle\langle 1| = |\mathrm{D}\rangle\langle\mathrm{D}|$ and $V_2 = |\mathrm{A}\rangle\langle\mathrm{A}|$. To increase the accuracy of the high temperature approximation (HTA) we include additional correction terms [9] for which we replace

$$
\begin{aligned}
\mathcal{L}_{\mathrm{ex}} \;\to\; & \mathcal{L}_{\mathrm{ex}} - \sum_{m=1}^{N}\frac{2\lambda}{\beta\hbar^2}\frac{2\nu}{\gamma_1^2-\nu^2}V_m^{\times}V_m^{\times} \\
\Theta_m \;\to\; & \Theta_m - \frac{2\lambda}{\beta\hbar}\frac{2\nu^2}{\gamma_1^2-\nu^2}V_m^{\times}.
\end{aligned}
\tag{S41}
$$

The hierarchy eq. (S41) can be truncated for a sufficiently deep hierarchy level $\sum_{m=1}^{N_{\mathrm{sites}}} n_m > N_{\mathrm{max}}$. Convergence can be tested by comparing deviations in the dynamics with increasing truncation level.

Simulations in the main text are run with the *QMaster* software package [10]. *QMaster* is a high-performance implementation of HEOM which rests on an efficient parallelization scheme taking advantage of the high computational throughput of modern hardware architectures, such as GPUs. *QMaster* has provided the necessary foundation for many practical applications ranging from detailed mechanistic studies for systems with structured spectral densities [7, 8] to simulations of large multi-protein light-harvesting architectures with more than 90 pigments [11]. All HEOM calculations presented here are run with $N_{\mathrm{max}} = 12$, with timesteps of less than 1 fs, and a temperature of 295 K for the thermal baths.

### B. HEOM: Shifted Drude-Lorentz vs underdamped Brownian oscillators

In section I, above, we relate the effective spectral density describing the high-frequency vibration and its thermal environment ($J_{\mathrm{eff}}(\omega)$) in the electronic basis to the vibrational environment ($J_{\mathrm{vib}}(\omega)$) in the vibronic basis. The interaction of the exciton system with the underdamped vibrational mode is described in the electronic basis by a Brownian Oscillator

$$
J_{\mathrm{elec}}^{\mathrm{BO}}(\omega) = 2\lambda\frac{2\gamma_{\mathrm{vib}}\Omega_{\mathrm{vib}}^2\omega}{4\gamma_{\mathrm{vib}}^2\omega^2+(\Omega_{\mathrm{vib}}^2-\omega^2)^2}.
\tag{S42}
$$

Exciton dynamics simulations are run in this work with the high-performance implementation of the HEOM approach provided by the *QMaster* software package. Instead of Brownian oscillators, *QMaster* represents structured spectral densities as a sum of shifted Drude-Lorentz peaks

$$
J_{\mathrm{elec}}^{\mathrm{DL}}(\omega) = \lambda\left(\frac{\gamma_{\mathrm{vib}}\omega}{\gamma_{\mathrm{vib}}^2+(\omega-\Omega_{\mathrm{vib}})^2}+\frac{\gamma_{\mathrm{vib}}\omega}{\gamma_{\mathrm{vib}}^2+(\omega+\Omega_{\mathrm{vib}})^2}\right).
\tag{S43}
$$

For small and intermediate values of the unitless parameter $\Gamma_{\mathrm{vib}}$, the Brownian Oscillator and the shifted Drude-Lorentz spectral density are nearly identical (see Fig. S2**a** and **b**). Deviations only occur for a stronger damping of the vibrational mode (see Fig. S2 **c**).

To assess to what extent the differences between $J_{\mathrm{elec}}^{\mathrm{DL}}(\omega)$ and $J_{\mathrm{elec}}^{\mathrm{BO}}(\omega)$ affect the exciton dynamics, we construct a superposition of shifted-Drude Lorentz peaks to mimic the Brownian Oscillator spectral density when $\Gamma_{\mathrm{vib}} = 100$ (the largest value considered in the main text). Deviations in the dynamics for this parameter value can be seen as

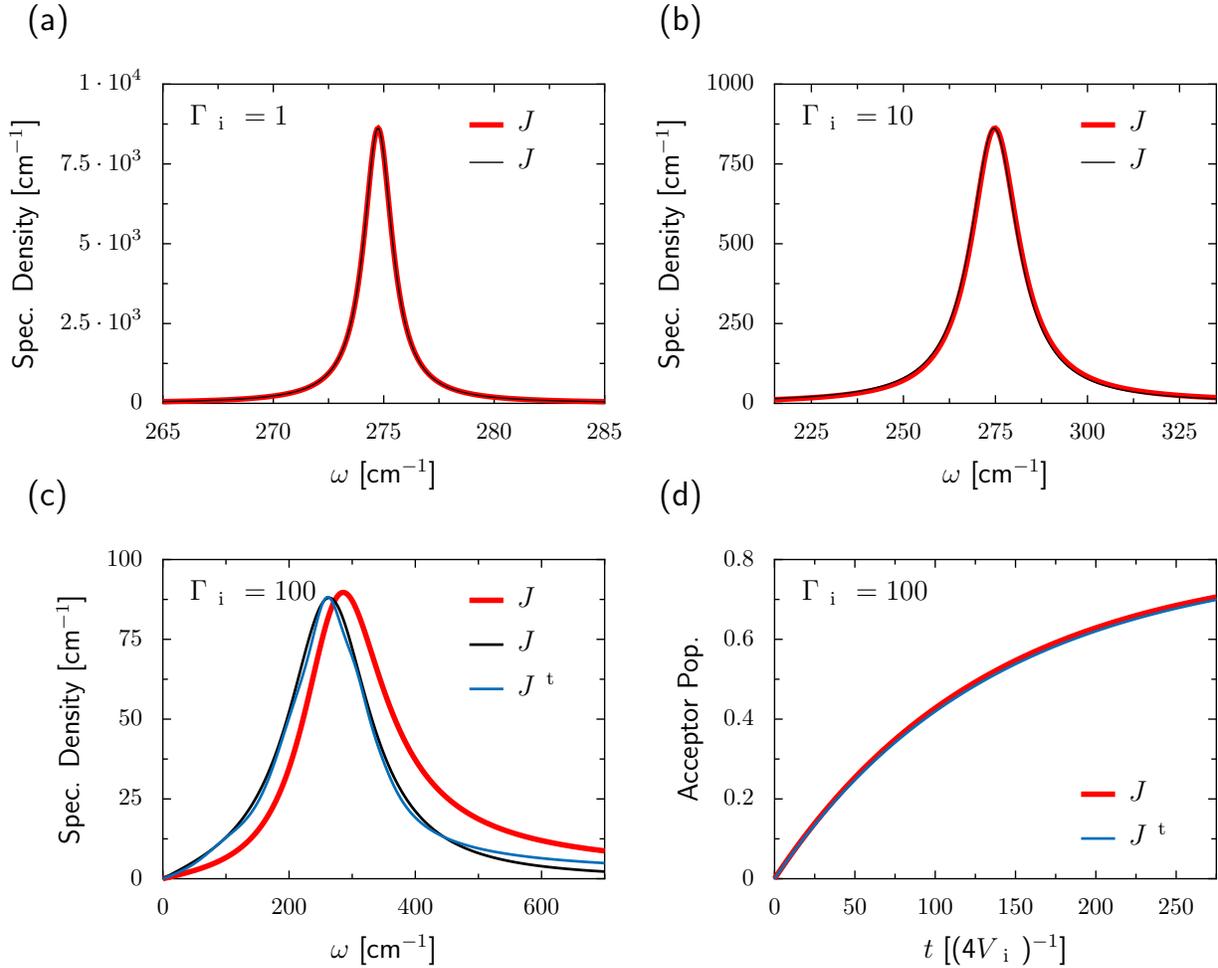

Figure S2: **The difference between Drude-Lorentz and Overdamped Brownian Oscillator spectral densities.** **(a)**–**(c)** comparison of the shifted Drude-Lorentz and Brownian oscillator spectral density for various values of $\Gamma_{\text{vib}}$. The remaining parameters are $\lambda = 24.7$, $\gamma_{\text{vib}}^{-1} = 67.4$ fs and $\hbar\Omega_{\text{vib}} = 274.7$ cm$^{-1}$. Additionally to the peak, all spectral densities exhibit a small continuous Ohmic Drude-Lorentz background (obtained from eq. (S43) by setting $\Omega = 0$) with $\lambda = 0.008$ cm$^{-1}$ and $\gamma^{-1} = 106.2$ fs. **(d)** population dynamics obtained with *QMaster* for the shifted Drude-Lorentz spectral density ($\Gamma_{\text{vib}}$=100) and the five peak fit $J_{\text{BO}}^{\text{fit}}$ (parameters are listed in Table S1).

an upper bound of the effect of using $J_{\text{elec}}^{\text{DL}}(\omega)$ instead of $J_{\text{elec}}^{\text{BO}}(\omega)$ in the course of this study. We approximate the corresponding Brownian Oscillator by a five-peak fit, illustrated in Fig S2**c**. The corresponding parameters are listed in Table S1. As is illustrated in Fig. S2**d**, we observe merely a minimal deviation in the exciton dynamics between $J_{\text{elec}}^{\text{DL}}(\omega)$ and $J_{\text{elec}}^{\text{BOfit}}(\omega)$. Therefore, we can conclude that the results presented in the main text are not affected by using shifted Drude-Lorentz peaks instead of underdamped Brownian Oscillators.

| peak | $\lambda_i$ [cm$^{-1}$] | $\gamma_i^{-1}$ [fs] | $\hbar\Omega_i$ [cm$^{-1}$] |
|---|---|---|---|
| peak 1 | 3.7 | 85 | 90 |
| peak 2 | 9.9 | 85 | 205 |
| peak 3 | 8.2 | 135 | 259 |
| peak 4 | 2.5 | 150 | 305 |
| peak 5 | 0.45 | 150 | 345 |

TABLE S1: Parameterization of the representation the five-peak representation $J_{\text{BO}}^{\text{fit}}$ of the Brownian Oscillator spectral density defined in eq. (S42) with $\lambda = 24.7$, $\gamma_{\text{vib}}^{-1} = 67.4$ fs and $\hbar\Omega_{\text{vib}} = 274.7$ cm$^{-1}$.

## C. HEOM: Bounds on the high-temperature approximation

The influence of the continuum of vibrational modes composing the environment coupled to the electronic states of the pigments enters into HEOM calculations through the energy-gap correlation function, $C(t)$:

$$\text{Re}[C(t)] = \frac{1}{2}(C(t) + C^*(t)) = \frac{1}{\pi}\int_0^\infty d\omega J(\omega)\coth[\beta\hbar\omega/2]\cos(\omega t) \tag{S44}$$

$$\text{Im}[C(t)] = \frac{1}{2}(C(t) - C^*(t)) = \frac{1}{\pi}\int_0^\infty d\omega J(\omega)\sin(\omega t) \tag{S45}$$

where $J(\omega)$ is the spectral density in the electronic basis (which can describe either an over- or under-damped Brownian oscillator). $C(t)$ can be expressed as a sum of exponential terms by solving the contour integration of eq. S44 over the upper half-plane of the complex $\omega$ axis. Because $\coth[\beta\hbar\omega/2]$ has an infinite number of poles the $\text{Re}[C(t)]$ includes a summation over the Matsubara frequencies $\nu_n = \frac{2\pi}{\beta}n$ for integer n> 0. The high-temperature approximation corresponds to ignoring the poles associated with the Matsubara frequencies and retaining only the poles associated with the spectral density.

The question before us is how to establish a regime of validity for the HTA without actually calculating dynamics in the presence of additional terms. We note that our purpose here is not to define all regimes where the HTA will be valid but rather to simply demonstrate that the current calculations fall into a regime where the HTA is valid. Following from the observation that the hyperbolic cotangent is well approximated by $\frac{1}{\beta\omega}$ when the $\beta\omega$ is small, the general rule of thumb has been to bound the regime of validity for the HTA by $\beta\gamma < 1$ for an overdamped Brownian oscillator. In order to extend this argument to the underdamped Brownian oscillator, however, we require a more generic approach. In the following we consider both the over- and under-damped Brownian oscillator spectral densities and describe a simple approach for bounding the error associated with the high-temperature approximation when considering a quickly reorganizing bath. In summary, we find that the high-temperature approximation is appropriate for the vibronic heterodimer studied in the main text when $\beta\gamma_{\text{elec}} < 1$, $\beta\Omega_{\text{vib}} < 3$, $\beta V_{\text{vib}} < 1$ and $\gamma_{\text{vib}} < \Omega_{\text{vib}}$. All of these bounds are satisfied for all calculations presented in the main text.

### 1. Overdamped Brownian Oscillators: $\beta\gamma_{\text{elec}} < 1$

In the main text an overdamped Brownian oscillator spectral density is used to describe the broad range of low-frequency vibrations that couple to the pigment excited states (the 'electronic' environment in the vibronic basis). The over-damped Brownian oscillator spectral density

$$J_{\text{OD}}(\omega) = 2\lambda_{\text{elec}}\frac{\omega\gamma_{\text{elec}}}{\omega^2 + \gamma_{\text{elec}}^2} \tag{S46}$$

has simple poles at $\omega = \pm i\gamma$, so eq. S44 can be solved by performing the contour integration over the upper half plane to give

$$\text{Re}[C(t)] = \gamma_{\text{elec}}\,\lambda_{\text{elec}}\,e^{-\gamma_{\text{elec}}t}\cot[\frac{\beta\gamma_{\text{elec}}}{2}] + \sum_{\nu_n}\frac{4\lambda_{\text{elec}}\gamma_{\text{elec}}}{\beta}\frac{\nu_n e^{-\nu_n t}}{\nu_n^2 - \gamma_{\text{elec}}^2} \tag{S47}$$

$$\nu_n = 2\pi n/\beta \tag{S48}$$

where the first term is kept in the HTA while the sum over the Matsubara frequencies is dropped.

We will start by assuming $\gamma_{\text{elec}} < \nu_1 = \frac{2\pi}{\beta}$ since after this the Matsubara contribution can go through divergences which would lead them to be the dominant term. In this limit it is important to notice that the Matsubara terms all decay more rapidly as a function of time ($e^{-\nu_n t}$) then the HTA term ($e^{-\gamma_{\text{elec}}t}$). Now we assume the bath is quickly reorganizing which implies all electronic dynamics of the system evolve more slowly than either $e^{-\gamma_{\text{elec}}t}$ or $e^{-\nu_n t}$. As a result, we can determine the relative importance of the terms neglected in the HTA by considering the sum of their integrated areas

$$A_{\text{HTA}} = \int_0^\infty \gamma_{\text{elec}}\,\lambda_{\text{elec}}\,e^{-\gamma_{\text{elec}}t}\cot[\frac{\beta\gamma_{\text{elec}}}{2}]dt = \lambda_{\text{elec}}\cot[\frac{\beta\gamma_{\text{elec}}}{2}] \tag{S49}$$

$$A_{\text{Mats}} = \sum_{\nu_n}\int_0^\infty\frac{4\lambda_{\text{elec}}\gamma_{\text{elec}}}{\beta}\frac{\nu_n e^{-\nu_n t}}{\nu_n^2 - \gamma_{\text{elec}}^2}dt = \frac{2\lambda_{\text{elec}}}{\beta\gamma_{\text{elec}}} - \lambda_{\text{elec}}\cot[\frac{\beta\gamma_{\text{elec}}}{2}]. \tag{S50}$$

By comparing the relative magnitudes of $A_{\text{HTA}}$ and $A_{\text{Mats}}$ we can determine the a bound on the regime where the correlation function can be well described without the Matsubara frequencies. The resulting ratio is given by

$$\frac{A_{\text{Mats}}}{A_{\text{HTA}}} = \frac{2}{\beta\gamma_{\text{elec}}}\tan(\frac{\beta\gamma}{2}) - 1 \tag{S51}$$

which can be solved numerically to determine that $> 90\%$ of the correlation function is captured by the HTA if $\beta\gamma < 1$, the same bound that has been described previously in the literature [9] for an overdamped Brownian oscillator.

### 2. Underdamped Brownian Oscillators: $\beta\Omega_{\text{vib}} < 3$

In our discussion of the overdamped Brownian oscillator we demonstrated that we can inspect the relative magnitude of the integrated contributions to the correlation function in order to establish the regime of validity for the HTA. Here we use this same approach to describe the bounds on the HTA for an underdamped Brownian oscillator, for which we are not aware of any previous descriptions for a simple rule of thumb analogous to what has been given for the overdamped Brownian oscillator.

The underdamped Brownian oscillator spectral density is given by

$$J_{\text{vib}}(\omega) = 2\lambda_{\text{vib}}\frac{2\gamma_{\text{vib}}\omega\Omega_{\text{vib}}^2}{(\Omega_{\text{vib}}^2 - \omega^2)^2 + 4\gamma_{\text{vib}}^2\omega^2}, \tag{S52}$$

and has four poles at $\omega = \pm\xi \pm i\gamma_{\text{vib}}$ where $\xi = \sqrt{\Omega_{\text{vib}}^2 - \gamma_{\text{vib}}^2}$. The real component of the corresponding correlation function (again solved by contour integration over the upper half-plane) is given by

$$\text{Re}[C(t)] = \frac{\lambda(\xi^2 + \gamma^2)}{\xi}e^{-t\gamma}\frac{\sin[\beta\gamma]\sin[\xi t] + \sinh[\beta\xi]\cos[\xi t]}{\cos[\beta\gamma] - \cosh[\beta\xi]} \tag{S53}$$

$$- \sum_n \frac{2\lambda\Omega^2}{\beta}\frac{4\gamma\nu_n e^{-\nu_n t}}{((\nu_n^2 + \Omega^2)^2 - 4\gamma^2\nu_n^2}$$

where the first line is retained in the HTA while the second line is the sum over Matsubara frequencies.

For the vibronic heterodimer discussed in the main text, the fastest timescale of transport is determined by the vibronic coupling $V_{\text{vib}}$ between the donor and the vibrationally excited acceptor state. It follows that the component of the correlation function arising from the Matsubara frequencies for the underdamped Brownian oscillators correspond to a Markovian correction (i.e. fast on the timescale of all other dynamics) as long as $\beta V_{\text{vib}} < 1$. Further, we will assume that the vibration is in the underdamped limit which implies $\xi \approx \Omega_{\text{vib}}$. Recall that for a vibronic heterodimer, $\Omega_{\text{vib}} \gg V_{\text{vib}}$ (if this was not true then transport would be dominantly electronic and not vibronic in character). It follows that the first quarter-oscillation in the HTA terms $(\sin[\xi t], \cos[\xi t], t \in [0, \frac{\pi}{2\xi}])$ occur on a timescale fast compared to transport. Thus the integrated area of the HTA correlation function from t=0 $\rightarrow \frac{\pi}{2\xi}$ corresponds to a lower-bound on its contribution to dynamics occurring on a timescale much faster than transport (i.e. the Markovian component of the correlation function). This provides the necessary comparison between the HTA and the Matsubara corrections in analogy to what we did for the overdamped Brownian oscillator,

$$A_{\text{HTA}} = \int_0^{\frac{\pi}{2\xi}} \frac{\lambda(\xi^2+\gamma^2)}{\xi}e^{-t\gamma}\frac{\sin[\beta\gamma]\sin[\xi t] + \sinh[\beta\xi]\cos[\xi t]}{\cos[\beta\gamma] - \cosh[\beta\xi]}dt = \frac{\lambda}{\xi}\frac{\xi\sin[\beta\gamma] + \gamma\sinh[\beta\xi]}{\cos[\beta\gamma] - \cosh[\beta\xi]} \tag{S54}$$

$$A_{\text{Mats}} = \sum_n \frac{2\lambda\Omega^2}{\beta}\frac{4\gamma\nu_n \int_0^\infty e^{-\nu_n t}dt}{((\nu_n^2 + \Omega^2)^2 - 4\gamma^2\nu_n^2} = 4\frac{\lambda\gamma}{\beta(\xi^2+\gamma^2)} - \frac{\lambda}{\xi}\frac{\xi\sin[\beta\gamma] + \gamma\sinh[\beta\xi]}{\cos[\beta\gamma] - \cosh[\beta\xi]} \tag{S55}$$

and thus we find

$$\frac{A_{\text{Mats}}}{A_{\text{HTA}}} = 1 - \frac{4\gamma\xi}{\beta(\xi^2+\gamma^2)}\frac{\cosh[\beta\xi] - \cos[\beta\gamma]}{\xi\sin[\beta\gamma] + \gamma\sinh[\beta\xi]}. \tag{S56}$$

The resulting ratio can be solved numerically to find that when $\beta\Omega_{\text{vib}} < 3$ the high-temperature approximation captures more than 90% of the correlation function, which is equivalent to $\beta\gamma_{\text{elec}} < 1$ found for the underdamped Brownian oscillator spectral density.

### D. Vibronic Redfield: Computational details

The master equation for the reduced density matrix is solved in the vibronic basis:

$$\frac{\partial \rho_{ij}}{\partial t} = -i\omega_{ij}\rho_{ij} - \sum_{kl} \mathcal{R}_{ijkl}\rho_{kl}. \tag{S57}$$

Here $\omega_{ij} = e_i - e_j$ is the energy difference of the two vibronic states and $\mathcal{R}_{ijkl}$ is the complex-valued Redfield relaxation tensor. The Redfield tensor is constructed from the correlation functions in the frequency domain, which are given by the spectral densities of the electronic and vibrational environments:

$$\mathrm{Re}\{C_{elec/vib}(\omega)\} = (n(\omega) + 1)\, J_{elec/vib}(\omega). \tag{S58}$$

Here $n(\omega) = \left(e^{\frac{\hbar\omega}{k_B T}} - 1\right)^{-1}$ is a thermal equilibrium population of modes at frequency $\omega$. The imaginary part of the Fourier-transformed correlation function can be obtained as [9]

$$\mathrm{Im}\{C_{elec/vib}(\omega)\} = \frac{1}{\pi}\wp \int_{-\infty}^{\infty} d\omega' \frac{\mathrm{Re}\{C_{elec/vib}(\omega')\}}{\omega - \omega'}. \tag{S59}$$

The $\wp$ denotes a principal value integral, which we calculate numerically. We note that the imaginary part of the Redfield tensor effectively renormalizes the vibronic state energies and is necessary to achieve correct excited-state equilibrium. The Redfield tensor can be also decomposed into the electronic and vibrational parts: $\mathcal{R}_{ijkl} = \mathcal{R}_{ijkl}^{elec} + \mathcal{R}_{ijkl}^{vib}$. Considering one explicit vibrational mode per site, the site basis states can be denoted as $|e_n^{\nu}\rangle = |g_1^{\nu_1}\rangle...|e_n^{\nu_n}\rangle...|g_N^{\nu_N}\rangle$, where $n$ denotes the electronic excitation and $\nu = (\nu_1,....,\nu_N)$ multiindex lists the vibrational states. The vibronic basis is given by the diagonalization of the system Hamiltonian $H_{sys}$ (S8), the transformation coefficients are $c_{n,\nu}^i$. To simplify the bookkeeping, we define the following auxiliary operators:

$$\mathcal{Y}_{ijkl}^{elec} = \sum_n \sum_{\nu\mu} c_{n,\nu}^i c_{n,\nu}^j c_{n,\mu}^k c_{n,\mu}^l C_{elec}(\omega_{ki})$$

$$\mathcal{Y}_{ijkl}^{vib} = \sum_{n,m,p} \sum_{\nu\mu\kappa\lambda} c_{n,\nu}^i c_{n,\mu}^j c_{m,\kappa}^k c_{m,\lambda}^l \left(\sqrt{\mu_p}\delta_{\nu_p\mu_p - 1} + \sqrt{\mu_p + 1}\delta_{\nu_p\mu_p + 1}\right)$$

$$\left(\sqrt{\lambda_p}\delta_{\kappa_p\lambda_p - 1} + \sqrt{\lambda_p + 1}\delta_{\kappa_p\lambda_p + 1}\right)\Pi_{q\neq p}\delta_{\nu_q\mu_q}\delta_{\kappa_q\lambda_q} C_{vib}(\omega_{ki}). \tag{S60}$$

The Redfield tensor is then expressed as [12]

$$\mathcal{R}_{ijkl}^{elec/vib} = -\left(\mathcal{Y}_{ijkl}^{elec/vib} + \mathcal{Y}_{jilk}^{elec/vib}\right) + \delta_{jl}\sum_s \mathcal{Y}_{siks}^{elec/vib} + \delta_{ik}\sum_s \mathcal{Y}_{sjls}^{elec/vib}. \tag{S61}$$

In this work we consider the vibrational mode present only on the acceptor. For the initial condition for the dynamics we consider an excitation on the donor, while the acceptor is in its ground state thermal equilibrium:

$$\rho_{ij}(0) = \sum_{\nu_A} c_{D,\nu_A}^i c_{D,\nu_A}^j e^{-\frac{\nu_A \Omega_{vib}}{k_B T}} \left(\sum_{\mu_A} e^{-\frac{\mu_A \Omega_{vib}}{k_B T}}\right)^{-1}. \tag{S62}$$

## III. DYNAMICS OF A VIBRONIC HETERODIMER

### A. Extracting overall rate from donor to acceptor from HEOM

Excitation dynamics in our model heterodimer can be complex, including coherent oscillations and multiple timescales of transport. The black lines in Fig. S3 show the acceptor population dynamics as a function of time

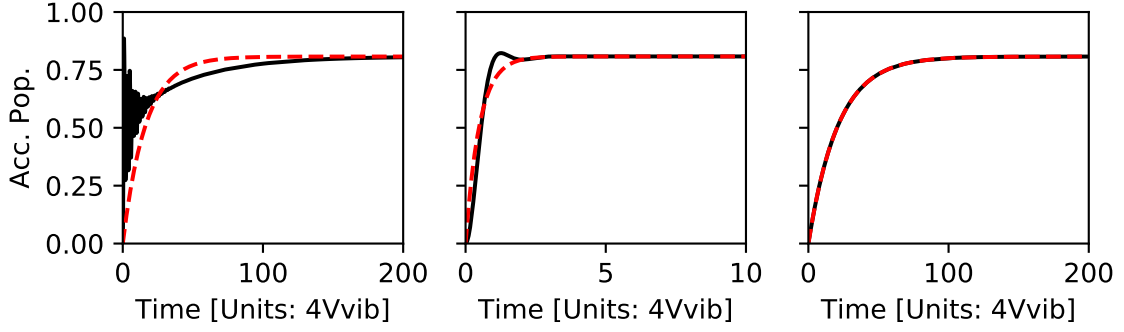

Figure S3: **Exponential fit to HEOM site population dynamics.** The acceptor site population dynamics simulated with HEOM (black lines) are plotted along with the best fit one-exponential curve (dashed red line). The HEOM simulation parameters are **(left)** $\Lambda_{\text{elec}} = 0.01$, $\Gamma_{\text{vib}} = 0.01$, **(center)** $\Lambda_{\text{elec}} = 0.01$, $\Gamma_{\text{vib}} = 1$, and **(right)** $\Lambda_{\text{elec}} = 0.01$, $\Gamma_{\text{vib}} = 100$.

following an initial donor excitation for $\Lambda_{\text{elec}} = 0.01$, $\Sigma_{\text{vib}} = 0$, and $\Gamma_{\text{vib}} = 0.01$ (left panel), 1 (middle panel), 100 (right panel). Clearly, of these simulations, only $\Gamma_{\text{vib}} = 100$ is completely described by a single timescale, the other two have some combination of multiple timescales and/or oscillitory features. Any complete description of the vibronic dynamics would, necessarily, be multi-dimensional - such as the regime diagram presented in Fig. 2 of the main text. To begin with, however, we use a simple single-exponential fit to describe the overall donor-to-acceptor transport rate ($k_{\text{a}\leftarrow\text{d}}$).

We use an analytic approach to determining the best-fit one-exponential timescale associated with the rise in acceptor population. We start with the rate equation for the donor population (eq.S63).

$$\frac{\mathrm{d}P_D(t)}{\mathrm{d}t} = -k_{\text{a}\leftarrow\text{d}}P_D(t) + k_{\text{d}\leftarrow\text{a}}P_A(t) \tag{S63}$$

Noting the equilibrium condition ($k_{\text{d}\leftarrow\text{a}} = k_{\text{a}\leftarrow\text{d}}\frac{P_D^{eq}}{P_A^{eq}}$) and the conservation of probability ($P_D(t) + P_A(t) = 1$), we can rewrite this equation in terms of only the time-dependent donor population (eq. S64).

$$\frac{\mathrm{d}P_D(t)}{\mathrm{d}t} = -k_{\text{a}\leftarrow\text{d}}[(1 + \frac{P_D^{eq}}{P_A^{eq}})P_D(t) - \frac{P_D^{eq}}{P_A^{eq}}] \tag{S64}$$

The solution to this equation (solved by transformation of variables) is a decaying exponential function (eq. S65). As a result, the rate of transport from the donor to the acceptor can be determined directly from the integral of the donor population dynamics, as shown in eq. S66.

$$P_D(t) = P_A^{eq} \exp[-\frac{k_{\text{a}\leftarrow\text{d}}}{P_A^{eq}}t] + P_D^{eq} \tag{S65}$$

$$\int_0^\infty [P_D(t) - P_D^{eq}]/(P_A^{eq})^2\mathrm{d}t = \frac{1}{k_{\text{a}\leftarrow\text{d}}} \tag{S66}$$

The resulting best-fit exponentials to the population dynamics of the acceptor are shown in dashed red curves (Fig. S3). As noted above, there are cases where a single timescale does a poor job describing the over transport dynamics, in particular, when the vibrational relaxation rate is small ($\Gamma_{\text{vib}} \ll 1$) a single exponential fit is a relatively poor description because there are two distinct timescales. We return to a more detailed analysis of the transport dynamics when considering the mechanism of transport, discussed below.

### B. Comparing HEOM and vibronic Redfield simulations

Vibronic Redfield theory is a perturbative treatment of exciton dynamics expressed in the vibronic basis. In order to establish the regime of validity for vibronic Redfield theory, we need to compare the resulting dynamics to

simulation using a numerically exact method, such as HEOM. Unfortunately, our HEOM simulations are expressed in the electronic basis; therefore, to compare vibronic Redfield and HEOM population dynamics we need to establish the connection between the electronic and vibronic parameters. This has been described in detail in section I, above. Here, we describe the resulting comparison of population dynamics and overall donor-to-acceptor transport rate ($k_{a \leftarrow d}$) across different parameter regimes.

Vibronic Redfield accurately describes exciton transport in the presence of Markovian thermal environments. In vibronic transport there are two thermal environments that must be considered: first an electronic environment that is coupled to the pigment excitation energies ($J_{\text{elec}}(\omega)$); second, there is a vibrational environment responsible for relaxation between states of the explicit vibration ($J_{\text{vib}}(\omega)$). When the high-frequency vibration is represented in the electronic basis by an underdamped Brownian oscillator peak in the spectral density, the corresponding vibrational environment is Ohmic ($J_{\text{vib}}(\omega) = \frac{\gamma_{\text{vib}}}{\Omega_{\text{vib}}}\omega$). As a result of $J_{\text{vib}}(\omega)$ being ohmic, the vibrational environment always relaxes on timescales much faster than any other dynamics (i.e. it is Markovian). The electronic environment, however, relaxes on a timescale set by the inverse peak-width ($\Gamma_{\text{elec}}^{-1}$). When $\Gamma_{\text{elec}}$ is very large, the electronic environment relaxes on timescales much faster than transport and can also be considered Markovian. In this case, we find that vibronic Redfield population dynamics reproduce HEOM simulations (Fig. S4). The corresponding donor-to-acceptor rates are shown in Fig. S5. As $\Gamma_{\text{elec}}$ decreases, however, the quality of the vibronic Redfield simulations also decrease with increasing $\Lambda_{\text{elec}}$, as seen by the poor reproduction of the donor-to-acceptor rates (Fig. S5).

Overall we find that the appropriateness of vibronic Redfield theory, like traditional Redfield, is system dependent. We can see that vibronic Redfield provides an accurate description of transport in the Markovian regimes simulated in the Main Text. Further, the unitless parameters $\Gamma_{\text{elec}}$ and $\Lambda_{\text{elec}}$ provide a simple tool to assess the applicability of vibronic Redfield to any specific molecular aggregate.

### C. Extracting effective transport and relaxation timescales

Excitation transport between states in the vibronic basis contains both vibronic transport within sub-blocks and vibrational relaxation between sub-blocks. In order to distinguish the relative timescales of these two different processes, we fit an effective rate matrix to the vibronic Redfield population dynamics upto the second vibrationally excited state ($\nu_{e/g} = 2$). This results in an unwieldy six states with a total of twelve rates composed of four vibronic transport rates and eight rates of vibrational transitions.

To simplify this problem, we make use of two basic principles. First, for each pair of forward/backward rates one of them can be determined from the other using detailed balance and knowledge of the equilibrium population distribution (eq. S67). Second, the rate of vibrational relaxation is, approximately, the same everywhere. The only differences in relaxation rate are the result of small changes in the energy gap between states arising from the perturbative correction to energies (discussed in section IC, above), which we will ignore in the following discussion. Together, these two principles reduce the problem to fitting six population curves to three free parameters making up the rate matrix (eq. S68). Note, that there is a different rate of vibronic transport within the second vibronic sub-block compared to the first, a consequence of the larger vibronic coupling. As a result, the final effective transport rate is reported as the average of transport rates over an thermal distribution of the donor states $|D, 0_g\rangle$ and $|D, 1_g\rangle$. Thermally averaged rates give equivalent results to weighting by a time-integrated fractional population.

$$
\begin{aligned}
k_{\text{relax}}^{(r)} &= k_{\text{relax}} \cdot \frac{P_{A,1_e}^{(eq)}}{P_{A,0_e}^{(eq)}} \\[2ex]
k_{\text{trans},1}^{(r)} &= k_{\text{trans},1} \cdot \frac{P_{D,1_g}^{(eq)}}{P_{A,0_e}^{(eq)}} \\[2ex]
k_{\text{trans},2}^{(r)} &= k_{\text{trans},2} \cdot \frac{P_{D,2_G}^{(eq)}}{P_{A,1_e}^{(eq)}}
\end{aligned}
\tag{S67}
$$

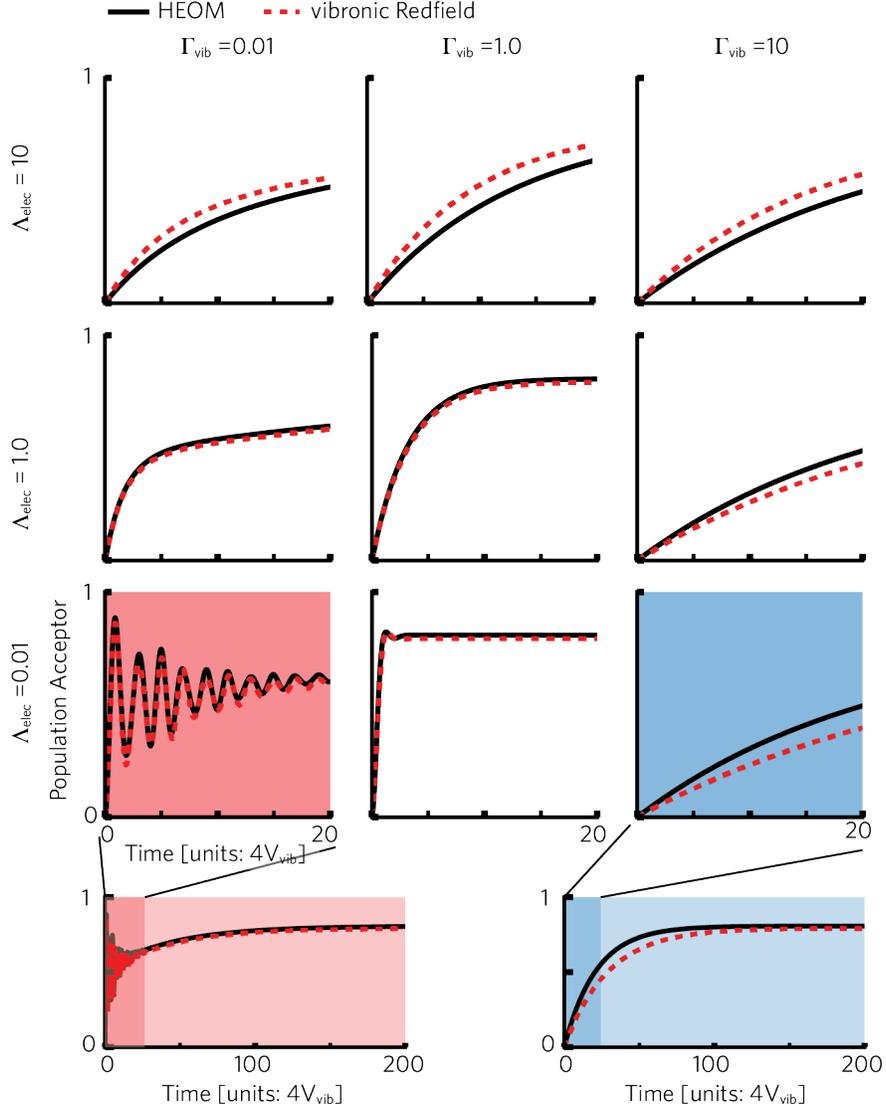

Figure S4: **Comparing HEOM and vibronic Redfield acceptor population dynamics.** The square grid of plots shows the short timescale acceptor population dynamics for both HEOM (black lines) and vibronic Redfield (dashed red lines) simulations. The two bottom plots show the long-timescale dynamics when $\Lambda_{elec} = 0.01$ and $\Gamma_{vib} = 0.01$ (left panel, red background) or $\Gamma_{vib} = 100$ (right panel, blue background).

$$
K = \begin{array}{c}
\begin{array}{cccccc}
|A, 0_e\rangle & |D, 0_g\rangle & |A, 1_e\rangle & |D, 1_g\rangle & |A, 2_e\rangle & |D, 2_g\rangle
\end{array} \\
\left[\begin{array}{cccccc}
-k_{relax}^{(r)} & 0 & k_{relax} & 0 & 0 & 0 \\
0 & -(k_{relax}^{(r)} + k_{trans,1}) & k_{trans,1} & k_{relax} & 0 & 0 \\
k_{relax}^{(r)} & k_{trans,1} & -(k_{trans,1}^{(r)} + k_{relax}^{(r)} + k_{relax}) & 0 & k_{relax} & 0 \\
0 & k_{relax}^{(r)} & 0 & -(k_{relax} + k_{trans,2}) & k_{trans,2}^{(r)} & k_{relax} \\
0 & 0 & k_{relax}^{(r)} & k_{trans,2} & -(k_{trans,2}^{(r)} + k_{relax}) & 0 \\
0 & 0 & 0 & k_{relax}^{(r)} & 0 & -k_{relax}
\end{array}\right]
\end{array} \quad \text{(S68)}
$$

In Figure 2 of the main text, we plot the contours which differentiate the 'transport limited' and 'relaxation limited' regimes in orange and green lines, respectively. We consider a vibronic dimer to show transport (relaxation) limited dynamics if $k_{trans}$ ($k_{relax}$) is the smaller of the two rates and no more than 0.04 [units of $4V_{vib}$] larger than the overall rate from donor to acceptor ($k_{a \leftarrow d}$). This provides a rudimentary separation between the two limiting regimes.

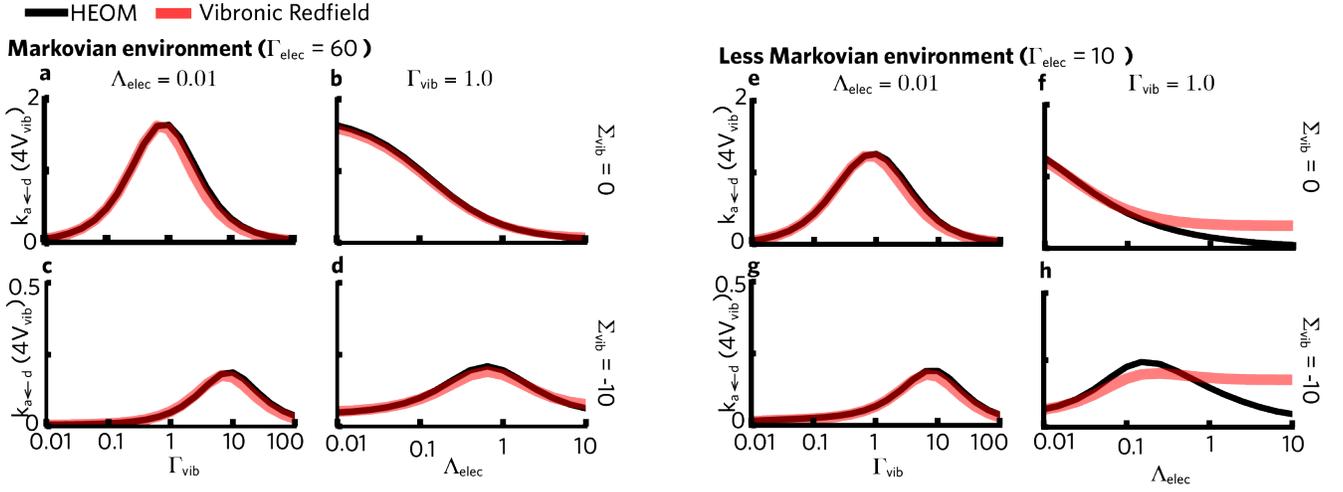

Figure S5: **Comparing $k_{a \leftarrow d}$ extracted from HEOM and vibronic Redfield simulations.** **(a-d)** Populations dynamics were simulated with the same parameters as Fig. 2 in the main text: $V_{vib} = 0.788$ cm$^{-1}$, $\gamma_{elec} = 50$ cm$^{-1}$, and $E_d$ - $E_a = 350$ cm$^{-1}$. **(e-h)** Populations dynamics were simulated with the same parameters as Fig. 2 in the main text, except that $\gamma_{elec} = 7.8$ cm$^{-1}$.

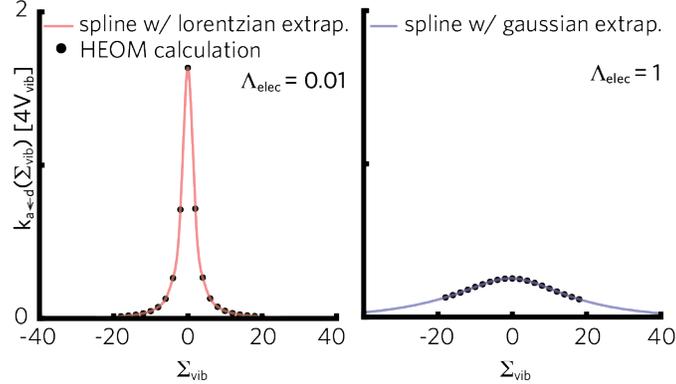

Figure S6: **Fitting $k_{a \leftarrow d}$ as a function of $\Sigma_{vib}$.** We extract $k_{a \leftarrow d}$ from HEOM simulations (black dots) where $\Gamma_{vib} = 1$ and $\Lambda_{elex} = 0.01$ (left panel) or $1.0$ (right panel). The red (blue) line is a spline fit with a Lorentzian (Gaussian) extrapolation.

### D. Disorder and ensemble distribution of $k_{a \leftarrow d}$

We simulate the probability distribution of the overall donor-to-acceptor rate over a random ensemble of energy gaps between the donor and vibronically excited accpetor states ($\Sigma_{vib}$) using a monte carlo simulation. The disorder between the donor and vibrationally excited acceptor state depends on the excitation energies of both pigments. The width of the disordered distribution in $\Sigma_{vib}$, therefore, should correspond to the convolution of the two inhomogeneous distributions and be larger than the width of either distribution alone. For simplicity, we randomly sample values of $\Sigma_{vib}$ from a Gaussian distribution with a standard deviation of 10 $V_{vib}$ which we estimate as a lower bound for a single site of a phycobiliprotein (Section IV D, below).

In each realization of $\Sigma_{vib}$ we calculate the overall rate from donor to acceptor ($k_{a \leftarrow d}$). We simulate 100,000 realizations and plot the resulting probability distribution in Fig. 4b of the main text. Because we need to run a large number of simulations, we cannot use HEOM (or even vibronic Redfield) to extract the overall donor-to-acceptor rates. Instead, having already calculated these rates for a range of $\Sigma_{vib}$ values using HEOM, we fit the resulting $k_{a \leftarrow d}(\Sigma_{vib})$ curve for both small ($\Lambda_{elec} = 0.01$) and large ($\Lambda_{elec} = 1$) thermal fluctuations (Fig. S6). In each case, for detuning values between -18 and 18 we use a cubic spline interpolation between HEOM data points. For data points outside of this region we extrapolate using a Lorentzian (eq. S69) or Gaussian (eq. S70) function for $\Lambda_{elec} = 0.01$ or $1.0$, respectively. The extrapolation parameters are fit to the last 3 data points on either side of 0. The resulting

functions and data points are plotted in Fig. S6. Note, for clarity and consistency, main text Figure 4a plots the fit $k_{a \leftarrow d}(\Sigma_{\mathrm{vib}})$ curve that is used to calculated the corresponding probability distributions in Fig. 4b.

$$L(\Sigma; A, \mu, \chi) = A \cdot \exp[-\chi \cdot |\Sigma - \mu|] \tag{S69}$$

$$G(\Sigma; A, \mu, \chi) = A \cdot \exp[-\chi^2(\Sigma - \mu)^2] \tag{S70}$$

## IV. EFFECTIVE DIMER MODELS FOR PIGMENT PROTEIN COMPLEXES

A variety of pigment protein complexes have been suggested to demonstrate coherent vibronic transport between energetically detuned pigments. Phycobiliproteins, with their flexible tetrapyrole (bilin) pigments, have been canonical examples of photosynthetic antennae that are thought to support coherent vibronic transport between pairs of pigments at room temperature [13–16]. In this section, we extract the Hamiltonian parameters from literature references that have suggested coherent vibronic transport in the phycobiliproteins (PC645, PE545, APC), and, as an example of non-dimer vibronic transport in photosynthetic antenna proteins, the Fenna-Mathews-Olson complex (FMO).

Our general strategy is to start with identifying the relevant pigments and vibrational mode. We then extract the reorganization energy of the low-frequency vibrational modes ($\lambda_{\mathrm{elec}}$), the Huang-Rhys factor of the high-frequency bridging vibration ($S = \frac{\lambda_{\mathrm{vib}}}{\Omega_{\mathrm{vib}}}$), the relaxation timescale of the high-frequency bridging vibration ($\gamma_{\mathrm{vib}}$), and the electronic coupling between pigments ($V$). We combine the interpigment coupling $V$ and the the Huang-Rhys factor (S) to determine the magnitude of the vibronic coupling ($V_{\mathrm{vib}} = V \sqrt{S} e^{-\frac{S}{2}}$). From these parameters we can determine the reduced parameters discussed in the main text: $\Lambda_{\mathrm{elec}} = \frac{\lambda_{deph}}{V_{vib}}$, and $\Gamma_{\mathrm{vib}} = \frac{\gamma_{vib}}{V_{\mathrm{vib}}}$.

### A. PC645

PC645 is a photosynthetic antenna complex of cryptophate algae. Coherent oscillations in 2d electronic spectroscopy of PC645 have been observed at room temperature [17]. In the center of PC645 there is a strongly coupled dimer of high-energy dihydro-biliverdins (DBV) pigments which rapidly transfer excitation energy to the lowest energy, peripheral phycocyanobilin 82 (PCB82) pigments. This transport was found to occur directly, without proceeding through intermediate energy pigments, despite the large energy gap (1600 cm$^{-1}$) and relatively weak electronic coupling ($V = 40$ cm$^{-1}$). Direct down-conversion from the DBVs to PCB82s was explained in terms of a coherent vibronic mechanism in Ref. [13], and here we extract the corresponding Hamiltonian parameters. A resonant vibrational mode ($\Omega_{vib} = 1580$ cm$^{-1}$) with a Huang-Rhys factor of 0.08 was suggested to mediate transport. The resulting vibronic coupling is $V_{vib} = 11$ cm$^{-1}$. The vibrational damping is $\gamma_{\mathrm{vib}} = \nu_2 = 15$ cm$^{-1}$ and the reorganization energy of the low-frequency modes is $\lambda_{\mathrm{elec}} = 187$ cm$^{-1}$. This gives $\Gamma_{\mathrm{vib}} = 1.4$ and $\Lambda_{\mathrm{elec}} = 17$.

We note that the donor state in PC645 is likely to be a delocalized exciton across the pair of high-energy DBVs. The presence of a delocalized donor state will decrease the effective reorganization energy (motional narrowing [18]). The magnitude of this effect, however, is at most a factor of 2 which would in the best case scenario for coherence give $\Lambda_{\mathrm{elec}} \approx 8$. A recent detailed computational study of PC645 suggests that $\Lambda_{\mathrm{elec}}$ in fact has a lower bound of 10 [19].

### B. PE545

PE545 is another biliprotein antenna complex found in cryptophyte algae. The protein structure is similar to PC645, though the constituent pigments are different. The core dimer is composed of a semi-symmetric pair of phycoerythrobillin 50 (PEB50). The energy gap between these two central pigments, however, are much larger than in PC645 ($\approx 1000$ cm$^{-1}$) and close to a prominent vibration with frequency $\Omega_{\mathrm{vib}} = 1111$ cm$^{-1}$. In the original model describing the spectroscopy by Novoderezhkin et al. [20] no coherent vibronic effects were included and the dynamics was described by (generalized) Förster theory. However, the central dimer was later proposed to be an example of vibronic transport [15, 16]. The parameters used when assigning a coherent vibronic transport mechanism include an electronic coupling of $V = 92$ cm$^{-1}$, vibrational Huang-Rhys factor of $S = 0.0578$, and vibrational damping of $\gamma_{\mathrm{vib}} = 5.3$ cm$^{-1}$. The low-frequency vibrational modes have a reorganization energy of $\lambda_{\mathrm{elec}} = 110$ cm$^{-1}$. The resulting vibronic coupling is $V_{vib} = 21.5$ cm$^{-1}$ and the reduced parameters are $\Gamma_{vib} = 0.25$ and $\Lambda_{\mathrm{elec}} = 5.1$.

### C. APC

Allophycoyanin is a biliprotein antenna complex found in the cyanobacterial phycobilisome. Womick and Moran offered the first specific proposal of coherent vibronic transport in photosynthesis to explain exciton transport between the phycocyanin $\alpha 84$ and $\beta 84$ sites [14]. The interpigment coupling is $V = 150$ cm$^{-1}$ and the resonant vibration was modeled as having a Huang-Rhys factor of $S = 0.125$ and an essentially infinite vibrational lifetime ($\gamma_{\text{vib}} = 0$). The low-frequency vibrations were suggested to have a reorganization energy of 329 cm$^{-1}$. The resulting vibronic coupling is $V_{\text{vib}} = 49.8$ cm$^{-1}$, which gives reduced vibronic parameters of $\Gamma_{\text{vib}} = 0$ and $\Lambda_{\text{elec}} = 6.6$.

### D. Energetic Disorder in Phycobiliproteins

It is hard to precisely determine the extent of energetic disorder arising from changes in pigment excitation energies in different protein conformations which exchange on timescales slow compared to exciton transport. It is expected that bilins experience a larger amplitude of energetic disorder compared to chlorophyll because of their relative flexibility. In PC645, temperature dependent fluorescence suggested an ensemble of excitation energies with a standard deviation of $\approx 150$ cm$^{-1}$ (the average of all of the site widths reported). This corresponds to a standard deviation of $13V_{\text{vib}}$. In PE545, the excitation energies have been reported to have a standard deviation of between 400 cm$^{-1}$ [20] and 500 cm$^{-1}$ [15], which corresponds to 18-22 $V_{\text{vib}}$. Overall we estimate the disorder associated with transport within a phycobiliprotein to be greater than 10 $V_{\text{vib}}$.

### E. FMO

Unlike the phycobiliproteins discussed above, the Fenna-Mathews-Olson complex (FMO) of green sulphur bacteria is not well described as an effective dimer for the purpose of understanding vibronic transport. However, the bounds on coherent vs incoherent transport described here still act as a simple rule-of-thumb to differentiate coherent from incoherent transport much like those rules that have been previously articulated for electronic transport [21, 22].

FMO was the first photosynthetic pigment protein complex to be suggested to have coherent dynamics based on oscillations in two-dimensional electronic spectroscopy. Effects of resonant vibrations in FMO have been studied by many groups [9, 23–25] though often with the aim of explaining features in the non-linear spectra. Here we isolate the Hamiltonian parameters used to describe FMO and assign a coherent vibronic transport mechanism by Nalbach et al. [26]. The energy gaps between the FMO chlorophyll pigments are on the order of couple of hundreds of cm$^{-1}$, so they considered an effective vibrational mode with frequency $\Omega_{\text{vib}} = 180$ cm$^{-1}$ and Huang-Rhys factor $S = 0.027$. The reorganization energy, obtained by integrating the low-frequency component of the spectral density, gives $\lambda_{\text{elec}} = 40$ cm$^{-1}$. The electronic coupling between pigments in FMO varies between 30-90 cm$^{-1}$. We thus find that $V_{\text{vib}} = 5$-15 cm$^{-1}$ and $\Lambda_{\text{elec}} = 2.7 - 8$. Since the high-frequency vibrations are un-damped, $\Gamma_{\text{vib}} = 0$ in this model.



\end{document}